\documentclass[11pt]{article}

\usepackage[final]{acl}

\usepackage{times}
\usepackage{latexsym}
\usepackage{booktabs}
\usepackage{pgfmath}
\usepackage[table]{xcolor}
\usepackage{collcell}
\usepackage{colortbl}
\usepackage{array}
\usepackage{amsmath}
\usepackage{float}
\usepackage{amsfonts}
\usepackage{enumitem}
\usepackage{placeins}
\usepackage{multirow}
\usepackage[most]{tcolorbox}
\usepackage{subcaption}
\usepackage{appendix}

\usepackage[T1]{fontenc}

\usepackage[utf8]{inputenc}

\usepackage{microtype}

\usepackage{inconsolata}

\usepackage{graphicx}

\title{The Magnitude Mirage: Rethinking Confidence for Reasoning-Intensive Retrieval}

\author{
  \textbf{ Jamie Holdcroft$^{1}$, Abdelrahman Abdallah$^{2}$, Adam Jatowt$^{2}$} \\
  $^{1}$UNSW Sydney \qquad $^{2}$University of Innsbruck \\
  \texttt{j.holdcroft@student.unsw.edu.au}\\
  \texttt{\{abdelrahman.abdallah,adam.jatowt\}@uibk.ac.at}
}

\begin{document}
\maketitle
\begin{abstract}
Many production RAG systems implement retrieval abstention by thresholding raw similarity scores, implicitly treating score magnitude as a confidence signal. We demonstrate that this practice degrades systematically as queries require reasoning beyond semantic matching. Across 11 retrieval architectures and 28 datasets, neural retrievers consistently assign high similarity scores to semantically related but constraint-violating documents, causing magnitude-based thresholds to collapse toward near-random abstention performance on logical and temporal reasoning tasks---a failure we term the Magnitude Mirage. To address this without computationally expensive alternatives, we conduct a large-scale empirical study of six zero-cost Query Performance Prediction (QPP) metrics across three cognitive tiers: semantic matching (BEIR), logical reasoning (BRIGHT), and temporal reasoning (TEMPO). Our central finding is that the key improvement comes from abandoning magnitude in favor of score-distribution signals: the gain from this shift exceeds the differences among distributional alternatives by a factor of 5–10$\times$. In particular, Score Gap ($s_1 - s_k$) and a practical adaptation of Score Magnitude and Variance (LSMV) improve abstention AUROC by up to 0.16 in settings where magnitude-based confidence provides little discriminative power. These methods require no additional inference, retraining, or latency, making them a practical zero-cost replacement for magnitude thresholding in deployed RAG systems.
\end{abstract}

\section{Introduction}

Retrieval-Augmented Generation (RAG) systems rely on external retrieval components to provide factual context to large language models \citep{lewis2020rag, guu2020realm, izacard2022atlas}. The reliability of such systems depends critically on their ability to detect retrieval failures. When the retrieved context does not contain relevant information, downstream generation models may hallucinate or produce incorrect answers \citep{asai2024selfrag, lee2025finetunerag, sun2024redeep}. Consequently, practical RAG pipelines require mechanisms that can predict retrieval success and abstain when confidence is low \citep{thakur2023knowing, asai2024selfrag}.

In many industrial RAG frameworks, this decision is often implemented using simple thresholding of raw retrieval scores, a design reflected in widely used toolkits such as LangChain and LlamaIndex \citep{langchain2024, llamaindex2024}. While computationally inexpensive, this approach implicitly assumes that absolute score magnitudes are comparable across queries and can reliably distinguish successful from unsuccessful retrieval.

\begin{figure*}[t]
\centering
\includegraphics[width=.9\textwidth]{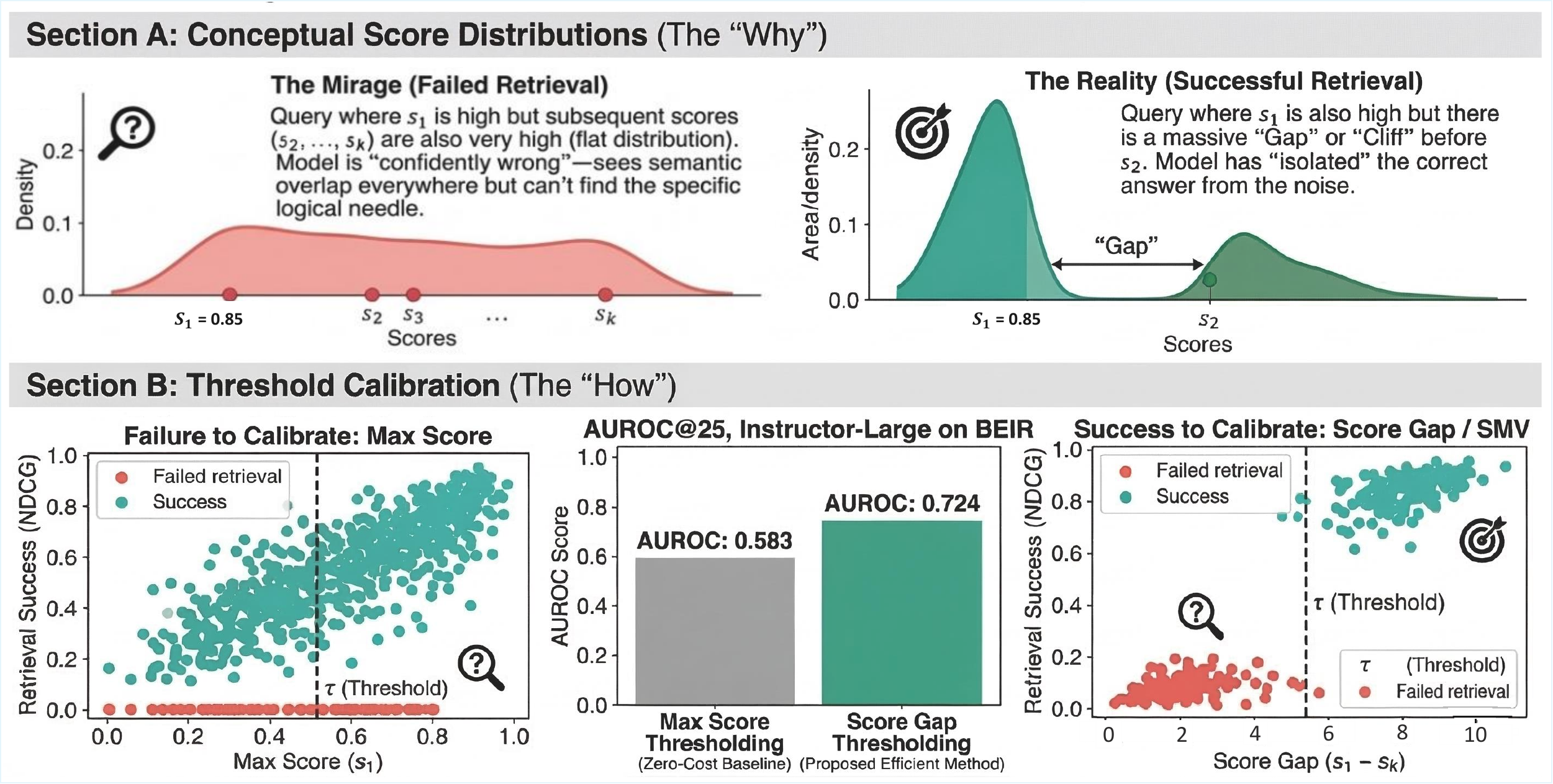}
\caption{The Magnitude Mirage vs.\ Reality. 
\textbf{Section A:} Two queries sharing identical top scores ($s_1 = 0.85$) 
exhibit fundamentally different score distributions. Failed retrieval produces 
a flat distribution; successful retrieval isolates the correct document via a 
large Score Gap. \textbf{Section B:} MaxScore thresholding cannot reliably distinguish 
successful from failed retrievals (AUROC: 0.583), while Score Gap thresholding 
achieves meaningful discrimination (AUROC: 0.724)---AUROC@25 averaged across the evaluated BEIR datasets for Instructor-Large.}
\label{fig:teaser}
\end{figure*}

Modern dense retrievers are typically trained to optimize semantic similarity between query and document embeddings \citep{karpukhin2020dpr, thakur2021beir}. Consequently, a retriever may assign a high similarity score to a document that matches the general semantics of a query while still violating key logical or temporal constraints necessary for correctness. In such cases, the raw retrieval score may appear highly confident even when the retrieved document is irrelevant to the true task requirement. We refer to this behavior as the \textit{Magnitude Mirage}: the misleading confidence conveyed by high absolute retrieval scores under reasoning-intensive retrieval constraints. As illustrated in Figure~\ref{fig:teaser}, two queries with identical maximum scores can exhibit fundamentally different score distributions. 

While recent work has explored LLM-based approaches for evaluating retrieved context, such methods introduce substantial latency and computational cost. This motivates revisiting classical Query Performance Prediction (QPP) techniques, which estimate retrieval success directly from the distribution of retrieval scores. 

In this work, we conduct a large-scale empirical study of zero-cost QPP methods in modern neural retrieval pipelines. To assess robustness across task complexity, we evaluate six QPP metrics across eleven retrieval architectures and twenty-eight datasets covering three cognitive tiers: semantic retrieval (BEIR) \citep{thakur2021beir}, logical reasoning (BRIGHT) \citep{su2025bright}, and temporal reasoning (TEMPO) \citep{abdallah2026tempo}. We measure predictive performance using AUROC for binary query success ($NDCG@k > 0$), which directly corresponds to the abstention decision in RAG systems. In open-domain QA and RAG pipelines, successful retrieval often requires only a single relevant passage within the retrieved context window to enable correct downstream generation \citep{lewis2020rag, karpukhin2020dpr}. We additionally report Pearson and Spearman correlations with $NDCG@k$ to evaluate continuous prediction quality.\\

\noindent Our contributions are as follows:

\begin{itemize}
    \item We provide a large-scale study of zero-cost QPP under reasoning-intensive retrieval, evaluating 6 metrics across 11 retrievers and 28 datasets spanning semantic, logical, and temporal retrieval.
    \item We demonstrate that magnitude-based confidence collapses systematically on reasoning-intensive benchmarks across all architectures tested, a regime largely unexplored in prior QPP work.
    \item We show that the primary gain comes from abandoning magnitude in favor of distributional estimation: all variance-based metrics substantially outperform MaxScore, while differing only marginally from one another.
    \item We provide end-to-end validation, showing that variance-based abstention improves downstream RAG correctness prediction relative to magnitude thresholding.
\end{itemize}

\section{Related Work and Background}

\subsection{Retrieval Reliability and RAG Abstention}

Irrelevant or misleading documents in a retrieved context can
substantially degrade downstream generation, even when only a single distracting passage is
included \citep{Cuconasu2024PowerofNoise}. While a growing body of work leverages generative models
to evaluate retrieved documents (\textit{LLM-as-a-judge}) \citep{Gu2024LLMAsAJudge,
liu2025LLMJudge}, these approaches incur substantial latency. Other methods improve
retrieval reliability through retriever alignment or reinforcement learning
\citep{zhou2025optimizing}, or cross-encoder reranking, but require retraining or additional
inference and do not provide lightweight abstention signals. In contrast, our work investigates
\textbf{zero-cost} confidence signals that operate directly on the retriever's score
distribution at inference time, requiring no additional model calls or retraining.

\subsection{Query Performance Prediction for Neural Retrieval}
Query Performance Prediction (QPP) estimates retrieval success without human relevance
judgments. QPP has been extensively studied in classical information retrieval \citep{tao2014smv, he2004qpp, carmel2010query}. Post-retrieval variance-based methods quantify the dispersion of top-ranked scores based on
the intuition that confident retrieval produces strong separation between relevant documents and
the background distribution \citep{he2004qpp, tao2014smv}.

Recent work has extended QPP to neural dense retrieval. \citet{10.1145/3765617}
show that classical QPP signals can differ substantially from sparse settings and that
magnitude-based signals often degrade in dense retrieval \citep{10.1145/3765617}.
Dense-specific adaptations include DenseQPP \citep{arabzadeh2021denseqpp}, QPP-PRP
\citep{10.1145/3765617}, and PDQPP \citep{datta2022pdqpp}; complementary work has explored
calibrating cosine similarity directly \citep{cosineadapter2024}. \citet{2024limitations} provide a cross-paradigm evaluation of QPP
robustness, finding that predictive power varies substantially across
collections and rankers and that QPP-driven selective query processing
yields only marginal gains. However, prior
evaluations largely focus on standard semantic benchmarks. We argue the limitation lies not merely in calibration, but in relying on magnitude itself as a confidence signal---a failure that becomes substantially more severe under reasoning-intensive retrieval, a regime prior QPP work has not examined.

\subsection{Reasoning-Intensive Retrieval}

Benchmarks such as BRIGHT and TEMPO introduce tasks requiring logical or temporal reasoning
beyond semantic similarity, where documents may be semantically close to a query while
violating its key constraints. How retrieval confidence signals behave in such
settings remains largely unexplored. Our work addresses this gap by evaluating QPP metrics
across semantic, logical, and temporal retrieval tasks.
\section{Methodology}
\subsection{Problem Formulation}
We formalize the standard Retrieval-Augmented Generation (RAG) first-stage retrieval process. Given a query $q$ and a corpus $C = \{d_1, d_2, \dots, d_N\}$, a retrieval model $R$ maps the query and corpus to a ranked list of the top-$k$ documents 
\[
D_k = [d_{(1)}, d_{(2)}, \dots, d_{(k)}],
\] 
where each retrieved document $d_{(i)}$ is associated with a retrieval score $s_i \in \mathbb{R}$, such that 
\[
s_1 \geq s_2 \geq \dots \geq s_k.
\]
Let $S_k = \{s_1, s_2, \dots, s_k\}$ denote the set of top-$k$ retrieval scores. 

Among the various classes of QPP methods, we focus specifically on \textit{zero-cost} QPP predictors that operate solely on the score distribution $S_k$. Formally, a QPP function
\[
f(S_k) \rightarrow c_q
\]
maps the score distribution to a scalar confidence estimate $c_q$ that approximates the true retrieval effectiveness of the query.

\subsection{The Magnitude Mirage}
An increasingly common heuristic for zero-cost RAG abstention relies on absolute magnitude thresholding. In practice, RAG frameworks such as LangChain and LlamaIndex apply fixed similarity thresholds to filter retrieved documents, passing only those where $s_i > \tau$ to the generator. This design implicitly assumes that raw retrieval scores are calibrated confidence signals and that a high score reliably indicates a relevant document regardless of query type. Under this assumption, the maximum score $s_1$ serves as a natural proxy for overall retrieval confidence: if the top-ranked document scores highly, the retrieval is considered successful.

As retrieval tasks shift from semantic matching to reasoning-oriented queries, the calibration of absolute retrieval scores degrades. As shown in Figure~\ref{fig:teaser}, even when two queries share an identical top score, MaxScore thresholding cannot reliably distinguish successful from failed retrieval --- achieving AUROC of only .583 compared to .724 for Score 
Gap on the same queries. This failure reflects a fundamental mismatch between what magnitude measures (semantic proximity) and what abstention requires (constraint satisfaction).

\subsection{Variance-Based QPP Metrics} To address the potential miscalibration of raw score magnitudes, we evaluate six zero-cost QPP metrics computed directly from the top-$k$
retrieval score distribution $S_k = \{s_1,\dots,s_k\}$. Distribution-based
metrics are motivated by the intuition that successful retrieval produces
greater separation between relevant documents and the background score
distribution.

\paragraph{Absolute Magnitude (Baseline):}  
\[MaxScore = s_1\]

\paragraph{Score Gap:} \[Gap@k = s_1 - s_k\]

\paragraph{Top-$k$ Standard Deviation:}

Let $\mu=\frac{1}{k}\sum_{i=1}^k s_i$. We define
\[
\sigma@k = \sqrt{\frac{1}{k} \sum_{i=1}^{k} (s_i - \mu)^2}
\]

\paragraph{Normalized Query Commitment (NQC):}
Adapting the classical NQC formulation of \citet{shtok2012querydrift} to dense retrieval, we normalize the score standard deviation by the mean absolute top-$k$ score:
\[
NQC@k = \frac{\sigma(S_k)}{\frac{1}{k}\sum_{i=1}^{k}|s_i|}
\]

\paragraph{Maximum Iterative Standard Deviation ($\sigma_{max}$)}\citep{Perez2010MaxStdDev}:\[\sigma_{max}@k = \max_{i \in [2, k]} \sigma(S_i)\]

\paragraph{Linearized Score Magnitude and Variance (LSMV):} A practical adaptation of SMV \citep{tao2014smv},  removing the logarithmic scaling used for sparse lexical retrieval, we evaluate the linearized dense-retrieval variant:
\[
LSMV@k = s_1 \times \sigma(S_k)
\]

\section{Experimental Setup}

\subsection{Datasets and Cognitive Tiers}

To evaluate the robustness of QPP signals across varying retrieval difficulty, we benchmark metrics across 28 datasets grouped into three cognitive tiers capturing different forms of reasoning constraints:

\begin{itemize}

\item \textbf{Semantic Matching (BEIR):} 
We utilize four representative datasets from the BEIR benchmark: SciFact, FiQA, NFCorpus, SciDocs. These datasets evaluate standard semantic retrieval tasks.

\item \textbf{Logical Reasoning (BRIGHT):} 
We evaluate on the BRIGHT benchmark, which introduces queries requiring logical constraints, multi-hop reasoning, and structured reasoning conditions across twelve datasets. In these tasks, simple semantic similarity is often insufficient to identify correct documents.

\item \textbf{Temporal Reasoning (TEMPO):} 
We include the TEMPO benchmark consisting of twelve datasets to evaluate retrieval under explicit and implicit temporal reasoning constraints. In this setting, documents may share high semantic similarity with a query while still violating required time conditions.
\end{itemize}

\subsection{Retrieval Models}

To ensure our findings generalize across retrieval paradigms, we evaluate QPP metrics across a diverse set of retrieval architectures.

\textbf{(1) Sparse Lexical Models:} 
We include BM25~\citep{robertson2009probabilistic} as a classical lexical retrieval baseline.
\textbf{(2) Dense Bi-Encoders:} 
We evaluate several widely used dense retrieval models, including BGE~\citep{xiao2024c}, E5~\citep{wang2022text}, Contriever~\citep{izacard2021unsupervised}, SBERT~\citep{reimers2019sentence}, SFR~\citep{meng2024sfrembedding}, Qwen~\citep{li2023generaltextembeddingsmultistage}, and Instructor-Large~\citep{su2023one}. \textbf{(3) Reasoning-Augmented Encoders:} 
We additionally include models explicitly trained or fine-tuned to handle structured or reasoning-intensive retrieval tasks, including Diver~\citep{long2025diver}, RaDeR~\citep{das2025rader}, and ReasonIR~\citep{shao2025reasonir}.

This diverse model suite enables evaluation across both classical and modern neural retrieval architectures.

\subsection{Evaluation Protocol}

For each query, we extract the raw retrieval score distribution $S_k$ from each model over the top-$k$ retrieved documents.

We evaluate QPP metrics across multiple context window sizes commonly used in RAG pipelines:
\[
k \in \{5, 10, 25, 50\}.
\]

\paragraph{Binary Retrieval Success}

We define retrieval success as the presence of at least one relevant document within the retrieved set:
\[
NDCG@k > 0.
\]
This criterion reflects the practical requirement that downstream generation typically depends on retrieving at least one relevant supporting passage within the context window \citep{lewis2020rag, karpukhin2020dpr}. Because this definition becomes less strict as $k$ increases, results at smaller cutoffs (e.g. $k=5$) serve as a stricter success criterion: retrieving a relevant document within only 5 candidates is substantially harder than within 25 or 50. Full results across all cutoffs are reported in Appendix Tables~\ref{tab:massive_appendix_beir}--\ref{tab:massive_appendix_tempo}; the relative ordering of QPP methods is consistent throughout.

\paragraph{Evaluation Metrics}

We evaluate the predictive quality of QPP signals using three complementary statistical measures: \textbf{ (1) AUROC (Area Under the Receiver Operating Characteristic Curve).}
This serves as our primary evaluation metric and simulates the RAG abstention decision. AUROC measures the ability of a QPP signal to distinguish between successful and failed retrieval instances across all possible decision thresholds, rather than relying on a fixed confidence threshold.  \textbf{(2) Pearson Correlation Coefficient ($r$).}
Measures the linear correlation between the predicted QPP confidence value and the observed retrieval effectiveness ($NDCG@k$). \textbf{(3)  Spearman Rank Correlation ($\rho$).}
Measures the monotonic relationship between predicted confidence and retrieval effectiveness, providing robustness to non-linear scaling effects. All metrics are computed per dataset and reported as macro-averages across datasets within each benchmark.

\subsection{Implementation Details}
Experiments were conducted on a compute node equipped with 4$\times$ NVIDIA H100-80GB GPUs, 512GB system memory, and AMD EPYC processors using PyTorch and HuggingFace Transformers. All dense models use FP16 inference with official Huggingface checkpoints. 
We release all code, evaluation scripts, dataset preprocessing pipelines, and experiment configurations to support full reproducibility.\footnote{\url{https://github.com/JamieHoldcroft/the-magnitude-mirage}}

\section{Results}

\subsection{The Magnitude Mirage Across Cognitive Tiers}

\begin{figure}[t]
    \centering
    \includegraphics[width=0.98\linewidth]{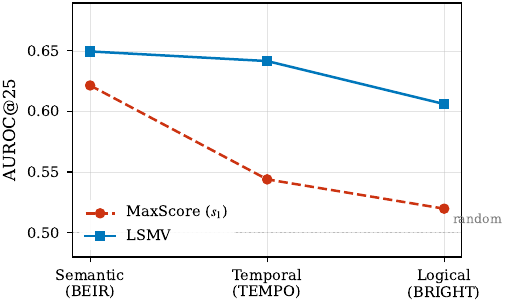}
    \caption{LSMV vs.\ MaxScore AUROC$@25$ illustrating systematic degradation of magnitude-based confidence as reasoning difficulty increases.}
    \label{fig:lineplot_reasonir}
\end{figure}

Table~\ref{tab:main_results_auroc} presents AUROC@25 scores for binary retrieval success across all three cognitive tiers. The results reveal a consistent pattern across all models: raw score magnitude (MaxScore) becomes an unreliable abstention signal for reasoning-intensive retrieval tasks, and its failure is systematic rather than incidental. Figure~\ref{fig:lineplot_reasonir} illustrates this collapse for ReasonIR: MaxScore AUROC degrades    steadily across cognitive tiers, approaching random performance on logical retrieval, while LSMV maintains stable discrimination.

On standard semantic retrieval (BEIR), MaxScore performs adequately for some models---BGE (.749) and SFR (.754) achieve reasonable discrimination---but already show instability for reasoning-augmented encoders such as ReasonIR (.621) and RaDeR (.666). The degradation becomes dramatic on reasoning-intensive tasks. On TEMPO, MaxScore collapses to near-random performance for Qwen (.536) and ReasonIR (.544). On BRIGHT, the failure is universal: \textit{every} model tested produces MaxScore AUROC values in the range .520--.611, providing essentially no actionable abstention signal. This confirms the core claim of the Magnitude Mirage: neural retrievers confidently assign high scores to logically or temporally incorrect documents, and the raw magnitude of those scores carries little reliable information about retrieval success.

\subsection{Variance-Based Metrics Restore Calibration}

In contrast, variance-based QPP signals---particularly Score Gap and LSMV---consistently recover meaningful abstention performance across all three tiers. On BEIR, Score Gap and LSMV improve over MaxScore for nearly every model, with gains of up to .158 AUROC (Instructor-Large: .583 $\to$ .741). On TEMPO, where MaxScore degrades most severely, Score Gap achieves gains of up to .125 AUROC (Qwen: .536 $\to$ .661). On BRIGHT, Score Gap and LSMV improve over MaxScore by margins ranging from .060 to .091 across dense models. These gains are consistent across both dense bi-encoders and reasoning-augmented encoders, suggesting that variance-based metrics capture a structural property of confident retrieval that persists across architectures: when a retriever correctly identifies a relevant document under complex constraints, that document statistically separates from the background distribution, producing a measurable score gap.


\begin{table*}[t]
\centering
\small
\setlength{\tabcolsep}{5pt}
\begin{tabular}{l >{\columncolor{gray!15}}c cc >{\columncolor{gray!15}}c cc >{\columncolor{gray!15}}c cc}
\toprule
\textbf{Model} & \multicolumn{3}{c}{\textbf{BEIR (Semantic)}} & \multicolumn{3}{c}{\textbf{TEMPO (Temporal)}} & \multicolumn{3}{c}{\textbf{BRIGHT (Logical)}} \\ 
\cmidrule(lr){2-4} \cmidrule(lr){5-7} \cmidrule(lr){8-10}
& Max & Gap & LSMV & Max & Gap & LSMV & Max & Gap & LSMV \\
\midrule
\textit{Sparse Models} & & & & & & & & & \\
BM25 & \textbf{.701} & .695 & .700 & .472 & \textbf{.526} & .510 & .611 & .622 & \textbf{.624} \\
\midrule
\textit{Dense Models} & & & & & & & & & \\
BGE & .749 & .805 & \textbf{.810} & .626 & .701 & \textbf{.705} & .538 & \textbf{.608} & \textbf{.608} \\
E5 & .736 & .777 & \textbf{.787} & .648 & .709 & \textbf{.718} & .552 & .628 & \textbf{.633} \\
Contriever & .683 & .727 & \textbf{.733} & .591 & .705 & \textbf{.707} & .560 & .624 & \textbf{.626} \\
SBERT & .718 & .766 & \textbf{.772} & .638 & .691 & \textbf{.720} & .525 & \textbf{.605} & .590 \\
SFR & .754 & .790 & \textbf{.802} & .649 & .671 & \textbf{.690} & .553 & \textbf{.623} & .622 \\
Qwen & .698 & .787 & \textbf{.788} & .536 & \textbf{.661} & .654 & .571 & \textbf{.653} & .650 \\
Instructor-Large & .583 & .724 & \textbf{.741} & .563 & .664 & \textbf{.669} & .534 & \textbf{.619} & .608 \\
\midrule
\textit{Reasoning Encoders} & & & & & & & & & \\
Diver-Retriever & .665 & .788 & \textbf{.789} & .582 & .666 & \textbf{.684} & .543 & .605 & \textbf{.606} \\
RaDeR & .666 & .740 & \textbf{.752} & .570 & .646 & \textbf{.665} & .560 & \textbf{.645} & .640 \\
ReasonIR & .621 & .626 & \textbf{.650} & .544 & .638 & \textbf{.642} & .520 & \textbf{.611} & .606 \\
\bottomrule
\end{tabular}
\caption{Comparison of zero-cost QPP metrics across three cognitive tiers. We report AUROC@25 scores for binary retrieval success ($NDCG@25 > 0$). Max = $s_1$, Gap = $s_1 - s_{25}$, and LSMV represents the Linearized Score Magnitude and Variance metric. The baseline magnitude column is shaded gray; best result per row--tier is \textbf{bolded}.}
\label{tab:main_results_auroc}
\end{table*}

\subsection{Comparison Across All Six QPP Metrics}

Full results for all six QPP metrics are provided in Appendix Table~\ref{tab:all_metrics_25}. First, all variance-based alternatives uniformly and substantially 
outperform MaxScore across all settings. Across dense models on BEIR, the improvement from switching to any variance-based metric ranges from .048 (SFR) to .158 (Instructor-Large), while the spread \textit{among} variance-based metrics for the same models is at most .017. On BRIGHT, where 
MaxScore AUROC is compressed into the narrow band .525--.571 
across dense models, variance-based metrics expand this to 
.590--.653---a recovery of meaningful discrimination from 
near-random baselines. The pattern is consistent: the gain from abandoning magnitude exceeds the differences among distributional alternatives by a factor of roughly $5$--$10\times$, confirming that the core contribution is the shift from magnitude to distributional estimation, rather than the specific choice of distributional statistic.

Second, Score Gap and LSMV emerge as the most consistent pair 
across tiers, though no single metric dominates universally. LSMV achieves the best or joint-best AUROC across the majority of BEIR configurations, while Score Gap provides stronger results on several BRIGHT and TEMPO settings. NQC offers complementary strength for sparse retrieval: BM25 on TEMPO shows NQC (.602) substantially outperforming Score Gap (.526), reflecting its stability under the heavier-tailed score distributions of term-frequency models. The key practical implication is that any variance-based signal provides more reliable abstention where magnitude fails---the specific choice is secondary.
Per-subset and per-cutoff breakdowns are provided in Appendix Section~\ref{sec:appendix_per_subset} and Tables~\ref{tab:massive_appendix_beir}--\ref{tab:massive_appendix_tempo} where the results remain consistent.

\subsection{Correlation with Continuous Retrieval Effectiveness}

Pearson and Spearman correlations with NDCG@$k$, as comprehensively reported in the Appendix tables, confirm the same ordering. MaxScore shows weak or near-zero linear correlation with retrieval effectiveness on reasoning-intensive tasks (e.g., Pearson $r < .03$ for Qwen and BM25 on TEMPO at $k=25$), while Score Gap and LSMV achieve correlations up to .40 on the same benchmarks. Spearman correlations exhibit the same pattern, confirming the relationship is monotonic rather than an artifact of linear scaling assumptions. For example, Score Gap achieves Pearson $r = .467$ vs.\ MaxScore $r = .328$ for Qwen on BEIR, $r = .401$ vs.\ $r = .193$ on BRIGHT and $r = .303$ vs $r = .028$ on TEMPO.

\subsection{End-to-End RAG Abstention}
\label{sec:end_to_end_rag}

\begin{figure}[H]
    \centering
    \includegraphics[width=\linewidth]{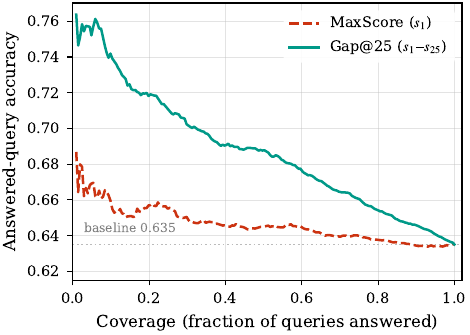}
    \caption{Coverage--accuracy trade-off macro-averaged across 14 
    retrievers on BRIGHT.}
    \label{fig:coverage_accuracy}
\end{figure}

We evaluate whether retrieval score geometry predicts downstream RAG
correctness across 11 retrievers and 19,376 BRIGHT query--retriever
pairs (11 retrievers $\times$ 1384 queries). For each query,
the top-5 retrieved documents are provided to \texttt{Qwen3.6-27B},
while GPT-4o independently judges answer correctness against BRIGHT
gold evidence documents to avoid self-evaluation bias. We simulate
abstention by ranking query instances by confidence and answering
only the top-$p\%$ of highest-confidence queries. Table~\ref{tab:rag_abstention} reports accuracy at representative coverage levels; Figure~\ref{fig:coverage_accuracy} shows the full continuous trade-off curve. Per-retriever AUROC with 95\% bootstrap confidence intervals and per-retriever operating points are reported in Appendix~\ref{sec:appendix_per_retriever_rag} (Tables~\ref{tab:per_retriever_rag_auroc}--\ref{tab:per_retriever_rag_operating}).

\begin{table}[H]
\centering
\small
\begin{tabular}{lccc}
\toprule
\textbf{Signal} & \textbf{@10\%} & \textbf{@25\%} & \textbf{@50\%} \\
\midrule
MaxScore ($s_1$)          & 65.8 & 65.6 & 64.5 \\
Gap@25 ($s_1 - s_{25}$)   & \textbf{74.1} & \textbf{71.0} & \textbf{68.8} \\
\midrule
Baseline (no abstention)  & \multicolumn{3}{c}{63.5} \\
\bottomrule
\end{tabular}
\caption{Answer accuracy (\%) at selective coverage levels, 
macro-averaged across 11 retrievers on 1,384 BRIGHT queries. 
All signals converge to the baseline at full coverage.}
\label{tab:rag_abstention}
\end{table}

As shown in Table~\ref{tab:rag_abstention}, MaxScore provides 
almost no advantage over the no-abstention baseline: even at 10\% 
coverage it reaches only 65.8\%, barely above the 63.5\% baseline, 
and for several retrievers falls below it.

Figure~\ref{fig:coverage_accuracy} shows the full coverage--accuracy
trade-off. MaxScore degrades rapidly toward baseline as coverage
increases, whereas Gap@25 maintains a consistent advantage
throughout the operating range. These results confirm that the
Magnitude Mirage propagates directly into downstream answer quality:
replacing $s_1$ thresholding with Gap@25 yields meaningfully higher
precision at selective operating points at zero additional cost. The per-retriever breakdown in Appendix~\ref{sec:appendix_per_retriever_rag} confirms this pattern is broadly consistent: all 11 retrievers show Gap@25 $\geq$ MaxScore AUROC for answer correctness, and three (\textit{Qwen2}, \textit{Nomic}, \textit{Contriever}) exhibit MaxScore AUROC \emph{below~0.5}; for \textit{Qwen2} the 95\% bootstrap CI [0.398,\,0.459] lies strictly below chance---raw similarity is \emph{anti-correlated} with answer success---while
\textit{Nomic} and \textit{Contriever} fall at the chance boundary
(CIs include 0.5). 

\section{Analysis and Discussion}

\subsection{Why Magnitude Fails in Reasoning Retrieval}

Modern dense retrievers are trained with contrastive objectives that 
reward semantic proximity, without explicitly encoding logical, 
mathematical, or temporal constraints. Consequently, the top-ranked 
score $s_1$ 
primarily reflects learned query–document compatibility and does not explicitly guarantee constraint satisfaction.
Multiple documents may achieve similarly high scores 
while violating critical task conditions, causing magnitude-based 
thresholding to fail to distinguish correct from incorrect retrieval. 
This explains the systematic MaxScore collapse observed on BRIGHT and TEMPO.

\subsection{Why Variance Signals Capture Retrieval Confidence}

\begin{figure}[t]
    \centering
    \includegraphics[width=\linewidth]{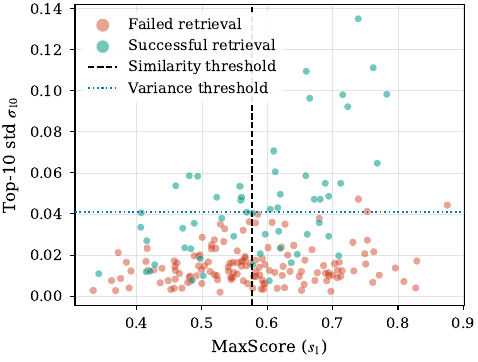}
    \caption{SBERT on BRIGHT TheoremQA.}
    \label{fig:scatter_sbert_theoremqa}
\end{figure}

Variance-based metrics operate on score distribution \textit{structure} 
rather than absolute magnitude. When a retriever correctly identifies 
a relevant document under complex constraints, that document often separates from the corpus background, producing a steeply declining score distribution with measurable gaps. As shown in Figure~\ref{fig:scatter_sbert_theoremqa}, on BRIGHT's 
TheoremQA subset, a vertical MaxScore threshold produces 
near-random class separation, while a horizontal Score Gap threshold partitions successful from failed retrievals.
This separation is not an artifact of corpus sparsity: reasoning-intensive benchmarks contain comparable or higher relevant document density per query than 
semantic retrieval tasks (BRIGHT: 4--5, TEMPO: 2--4, vs. BEIR: 1--5), suggesting that the 
observed score separation reflects retrieval confidence rather than a statistical property of the relevance 
annotations. Two observations argue against a benchmark-specific confound.  First, MaxScore degradation is consistent across both BRIGHT and TEMPO, which differ substantially in reasoning type, annotation methodology, and corpus construction. Second, the pattern holds across all 
11 retrieval architectures and all four cutoffs tested. Per-subset results in Appendix Section~\ref{sec:appendix_per_subset} 
further confirm this: Gap@25 outperforms MaxScore across the large 
majority of subsets in both BRIGHT and TEMPO independently.
Failed retrieval produces flatter distributions where many semantically similar documents receive comparable scores. This structural separation is architecture-agnostic: it emerges consistently across dense bi-encoders and reasoning-augmented encoders alike, explaining the uniform gains of Score Gap and LSMV.

\subsection{Implications for RAG System Design}

Our results suggest a simple revision to standard RAG abstention: replace absolute magnitude thresholding with a variance-based confidence signal. 
Across dense retrieval architectures and all three cognitive tiers, variance-based signals 
consistently outperform MaxScore, with improvements larger than the differences among the variance metrics themselves. This indicates that the primary benefit comes from abandoning magnitude-based confidence estimation rather than from the specific choice of variance metric. Because variance-based QPP metrics produce scalar confidence estimates analogous to existing similarity thresholds, they can directly replace magnitude-based filtering in deployed RAG pipelines without modifying retrieval or generation components. Score Gap and LSMV provide strong default choices given their consistent performance. Because these metrics operate solely on the top-$k$ score distribution already produced during retrieval, they can be integrated into existing RAG pipelines at \textit{zero cost}, without additional model calls, retraining, or latency. In practice, abstention thresholds can be calibrated on a held-out validation set to achieve a desired operating point, such as target answer precision or abstention rate.

\section{Conclusion}

We characterize the \textit{Magnitude Mirage}---the systematic failure of absolute retrieval score thresholds as confidence signals under logical and temporal reasoning constraints---demonstrating that this behavior consistently affects neural retrievers across diverse architectures. Our findings indicate that variance-based QPP metrics, specifically Score Gap and adaptations like LSMV, restore meaningful, albeit moderate, abstention signals at zero additional cost by measuring score distribution structure rather than raw magnitude. 

Our broader finding is that the shift from magnitude to distributional estimation matters more than which distributional statistic is chosen---the distribution-based metrics substantially outperform MaxScore. This suggests that the RAG community's  reliance on magnitude thresholding is not a calibration problem to be tuned, but a structural mismatch between semantic retrieval objectives and the requirements for constraint-based abstention. As retrieval tasks continue to grow in reasoning complexity, this mismatch will likely widen, making distribution-aware confidence estimation an increasingly critical component of reliable RAG system design.


\section*{Limitations}

First, while our end-to-end experiment demonstrates that variance-based 
abstention improves downstream answer quality on BRIGHT, this evaluation 
uses a single generator (\texttt{Qwen3.6-27B}) and judge (GPT-4o). 
Whether these gains generalize across generator architectures, prompting 
strategies, or answer evaluation protocols remains an open question.

Second, our experiments are limited to publicly available benchmarks. 
Real-world retrieval systems may exhibit different score calibration 
properties due to domain-specific corpora, proprietary query 
distributions, or custom retriever training procedures not represented 
here.

Third, we define binary retrieval success as $NDCG@k > 0$, reflecting 
the presence of at least one relevant document. For queries with a single 
gold document---common in both BRIGHT and TEMPO---this is the only 
meaningful binary criterion, as stricter grade thresholds collapse to the 
same condition. Results at $k=5$ serve as a naturally stricter operating 
point; the consistent ordering of QPP signals across all cutoffs in the 
appendix tables confirms that our conclusions are robust to this choice.

Finally, while we establish that variance-based signals substantially 
outperform magnitude thresholding, we do not characterize the absolute 
performance ceiling for zero-cost abstention. Whether the remaining gap 
relative to computationally expensive approaches such as cross-encoder 
reranking or LLM-based verification can be closed without additional 
inference remains an open question.

\bibliography{custom}

\appendix

\clearpage
\appendixpage            
\addappheadtotoc         
\numberwithin{figure}{section}
\numberwithin{table}{section}

This appendix provides supplementary material for \textit{The Magnitude Mirage}. It is organized as follows:

\begin{itemize}
\item Appendix~\ref{sec:appendix_comprehensive} provides comprehensive QPP evaluation results that complement the headline numbers reported in the main paper, including all six metrics at $k=25$, Pearson correlations across all benchmarks and cutoffs, and full per-benchmark tables at $k \in \{5,10,25,50\}$ for BEIR, BRIGHT, and TEMPO.
\item Appendix~\ref{sec:appendix_per_subset} reports per-subset AUROC@25 for MaxScore and Score Gap across every individual BEIR (4), BRIGHT (12), and TEMPO (12) subset, addressing reviewer requests for breakdowns of where the variance-based gains concentrate.
\item Appendix~\ref{sec:appendix_per_retriever_rag} provides the per-retriever breakdown of the end-to-end RAG abstention experiment (Section~\ref{sec:end_to_end_rag}) with 95\% bootstrap confidence intervals across 11 retrievers, and per-retriever selective-prediction operating points at coverage levels 10/25/50\%.
\item Appendix~\ref{sec:appendix_prompts} reproduces verbatim the two prompt templates used in the end-to-end RAG experiment (generator and judge), with full decoding parameters and parsing rules to support exact reproduction.
\end{itemize}

\paragraph{Key Tables}
\begin{itemize}
\item Table~\ref{tab:all_metrics_25}: AUROC@25 for all six zero-cost QPP metrics (MaxScore, Score Gap, Top-$k$ Std, NQC, Maximum Iterative Std, LSMV) across BEIR, BRIGHT, and TEMPO.
\item Table~\ref{tab:pearson_qpp_correlations}: Pearson correlations between MaxScore / Score Gap and $NDCG@k$ for every model, benchmark, and cutoff $k \in \{5,10,25,50\}$.
\item Tables~\ref{tab:massive_appendix_beir}, \ref{tab:massive_appendix_bright}, \ref{tab:massive_appendix_tempo}: full six-metric evaluation jointly reporting AUROC, Pearson $r$, and Spearman $\rho$ for BEIR, BRIGHT, and TEMPO at all four cutoffs.
\item Tables~\ref{tab:per_subset_beir}, \ref{tab:per_subset_bright}, \ref{tab:per_subset_tempo}: per-subset AUROC@25 for MaxScore and Gap@25 with the per-subset $\Delta$, macro-averaged across 11 retrievers.
\item Table~\ref{tab:per_retriever_rag_auroc}: per-retriever AUROC for downstream answer correctness in the end-to-end RAG experiment, with 95\% bootstrap CIs ($N{=}1000$).
\item Table~\ref{tab:per_retriever_rag_operating}: per-retriever selective-prediction answered-accuracy at abstention coverages 10/25/50\%, highlighting the 38-point Qwen2 coverage swing.
\end{itemize}

\paragraph{Key Figures}
\begin{itemize}
\item Figure~\ref{fig:prompt-generator}: generator prompt template used by \texttt{Qwen3.6-27B} (top-5 retrieved BRIGHT documents as context).
\item Figure~\ref{fig:prompt-judge}: judge prompt template used by GPT-4o (gold BRIGHT evidence as reference), deliberately chosen from a different model family to avoid self-evaluation bias.
\end{itemize}

\section{Comprehensive Per-Benchmark Results}
\label{sec:appendix_comprehensive}

This appendix collects the comprehensive evaluation tables behind the headline numbers in the main paper. Subsection~\ref{sec:appendix_all_metrics_at_25} reports all six QPP metrics at the headline cutoff $k=25$; Subsection~\ref{sec:appendix_pearson_overview} reports Pearson correlations across all four cutoffs; Subsection~\ref{sec:appendix_full_per_benchmark} provides the full per-benchmark tables at $k \in \{5,10,25,50\}$.

\subsection{All Six QPP Metrics at $k=25$}
\label{sec:appendix_all_metrics_at_25}

Table~\ref{tab:all_metrics_25} expands the three-metric main-paper table (MaxScore, Gap, LSMV) to all six zero-cost QPP signals (additionally: Top-$k$ Std, NQC, Maximum Iterative Std) at $k=25$, broken down by cognitive tier. The key observation is that every variance-based metric beats MaxScore on every dense retriever on BRIGHT and TEMPO. The \emph{spread} among variance-based metrics is small (typically $\leq 0.02$ AUROC) relative to the gap between any of them and MaxScore (often $0.05$--$0.16$), confirming that the operative axis of improvement is magnitude $\to$ distribution rather than the choice of specific distributional statistic.

\subsection{Pearson Correlations Across Benchmarks and Cutoffs}
\label{sec:appendix_pearson_overview}

Table~\ref{tab:pearson_qpp_correlations} provides a compact cross-benchmark view of the continuous prediction quality of MaxScore versus Score Gap, reporting Pearson $r$ between each signal and $NDCG@k$ for every model, benchmark, and cutoff $k \in \{5,10,25,50\}$. MaxScore correlations are near-zero or weak on reasoning-intensive benchmarks (e.g., $r \approx 0$ for Qwen and BM25 on TEMPO at most cutoffs), while Score Gap reaches $r=.40$+ on the same benchmarks. The pattern is monotone across cutoffs, confirming that the distinction is not an artifact of any specific choice of $k$.

\subsection{Full Per-Benchmark Tables at All Cutoffs}
\label{sec:appendix_full_per_benchmark}

Tables~\ref{tab:massive_appendix_beir}, \ref{tab:massive_appendix_bright}, and~\ref{tab:massive_appendix_tempo} provide the full six-metric evaluation across all $k \in \{5, 10, 25, 50\}$ for BEIR, BRIGHT, and TEMPO respectively, jointly reporting AUROC, Pearson ($r$), and Spearman ($\rho$). The relative ordering of QPP metrics is stable across all cutoffs, confirming that conclusions drawn at $k=25$ generalize across context window sizes.

\section{Per-Subset Results}
\label{sec:appendix_per_subset}

Tables~\ref{tab:per_subset_beir}, \ref{tab:per_subset_bright},
and~\ref{tab:per_subset_tempo} report AUROC@25 for MaxScore and Score
Gap at the subset level, macro-averaged across all 11 retrievers.
Gap@25 outperforms MaxScore on 3 of 4 BEIR subsets, 10 of 12 BRIGHT
subsets, and 10 of 12 TEMPO subsets. Exceptions are limited to a small
number of outlier subsets where corpus structure or query distribution
gives magnitude residual discriminative value; in all cases the gap is
small ($|\Delta| \leq 0.055$). These results confirm that the gains
reported in the main paper are broadly consistent across individual
datasets rather than driven by a small number of outlier subsets.

\section{Per-Retriever End-to-End RAG Abstention with Bootstrap CIs}
\label{sec:appendix_per_retriever_rag}

This appendix provides the per-retriever breakdown of the end-to-end RAG abstention experiment summarized in Section~\ref{sec:end_to_end_rag} (Table~\ref{tab:rag_abstention}, Figure~\ref{fig:coverage_accuracy}). Confidence intervals are computed by bootstrap resampling of the per-query (signal, judgment) pairs with $N{=}1000$ resamples.

Table~\ref{tab:per_retriever_rag_auroc} reports per-retriever AUROC for predicting downstream answer correctness, using each of the three confidence signals (MaxScore $s_1$, Score Gap at the RAG context window $s_1 - s_5$, and the paper-headline Score Gap $s_1 - s_{25}$). \textbf{All 11 retrievers exhibit Gap@25 AUROC $\geq$ MaxScore AUROC}, with non-overlapping 95\% CIs for the majority of dense retrievers. For three retrievers (\textit{Contriever}, \textit{Nomic}, and most strikingly \textit{Qwen2}) the MaxScore AUROC lies \emph{below 0.5}; for \textit{Qwen2} the CI [0.398,0.459] lies strictly below chance---raw similarity is statistically \emph{anti-correlated} with downstream answer success---while \textit{Nomic} and \textit{Contriever} fall at the chance boundary (CIs include 0.5). These cases provide the cleanest single-retriever evidence of the Magnitude Mirage in the entire study, and the Gap@25 signal recovers AUROC of 0.55--0.60 with statistical significance.

Table~\ref{tab:per_retriever_rag_operating} reports selective-prediction answered-accuracy at three abstention coverage levels (10\%, 25\%, 50\%) per retriever. The most striking case is \textit{Qwen2} at coverage 10\%: MaxScore-based abstention retains queries with only 39.1\% answered-accuracy---26 points \emph{below} the no-abstention baseline of 65.8\%---while Gap@25 retains queries with 76.8\% accuracy, an 11-point \emph{improvement} over baseline. The same set of queries, the same retriever, but a 38-point swing in downstream answer quality depending on which confidence signal is used for abstention. The macro-averaged $\Delta$ of +8.3 points at 10\% coverage (Table~\ref{tab:per_retriever_rag_operating}, last row) matches the macro figure in Section~\ref{sec:end_to_end_rag}.


\begin{table*}[t]
\centering
\small
\setlength{\tabcolsep}{4pt}
\begin{tabular}{l c c c c}
\toprule
\textbf{Retriever} & $N$ & \textbf{MaxScore} & \textbf{Gap@5 (ctx)} & \textbf{Gap@25} \\
\midrule
BM25 & 1384 & 0.573\,\textsuperscript{[0.542,\,0.603]} & 0.565\,\textsuperscript{[0.535,\,0.598]} & \textbf{0.586\,\textsuperscript{[0.554,\,0.618]}} \\
SBERT & 1384 & 0.517\,\textsuperscript{[0.486,\,0.548]} & 0.556\,\textsuperscript{[0.525,\,0.589]} & \textbf{0.562\,\textsuperscript{[0.530,\,0.593]}} \\
Contriever & 1384 & 0.488\,\textsuperscript{[0.457,\,0.519]} & 0.551\,\textsuperscript{[0.522,\,0.583]} & \textbf{0.558\,\textsuperscript{[0.531,\,0.591]}} \\
E5 & 1384 & 0.510\,\textsuperscript{[0.479,\,0.542]} & 0.524\,\textsuperscript{[0.493,\,0.557]} & \textbf{0.552\,\textsuperscript{[0.521,\,0.584]}} \\
BGE & 1384 & 0.514\,\textsuperscript{[0.483,\,0.545]} & 0.571\,\textsuperscript{[0.542,\,0.602]} & \textbf{0.577\,\textsuperscript{[0.546,\,0.606]}} \\
SFR & 1384 & 0.511\,\textsuperscript{[0.480,\,0.543]} & 0.552\,\textsuperscript{[0.519,\,0.584]} & \textbf{0.576\,\textsuperscript{[0.543,\,0.607]}} \\
Instructor-L & 1384 & 0.513\,\textsuperscript{[0.482,\,0.543]} & 0.555\,\textsuperscript{[0.525,\,0.585]} & \textbf{0.574\,\textsuperscript{[0.547,\,0.604]}} \\
Instructor-XL & 1384 & 0.525\,\textsuperscript{[0.490,\,0.557]} & 0.568\,\textsuperscript{[0.535,\,0.598]} & \textbf{0.576\,\textsuperscript{[0.546,\,0.607]}} \\
Qwen & 1384 & 0.528\,\textsuperscript{[0.497,\,0.558]} & 0.539\,\textsuperscript{[0.504,\,0.569]} & \textbf{0.551\,\textsuperscript{[0.520,\,0.580]}} \\
Qwen2 & 1384 & 0.429\,\textsuperscript{[0.398,\,0.459]} & 0.558\,\textsuperscript{[0.526,\,0.588]} & \textbf{0.585\,\textsuperscript{[0.551,\,0.615]}} \\
Nomic & 1384 & 0.481\,\textsuperscript{[0.450,\,0.512]} & 0.594\,\textsuperscript{[0.565,\,0.626]} & \textbf{0.603\,\textsuperscript{[0.572,\,0.634]}} \\
RaDeR & 1384 & 0.537\,\textsuperscript{[0.507,\,0.570]} & 0.524\,\textsuperscript{[0.490,\,0.557]} & \textbf{0.538}\,\textsuperscript{[0.503,\,0.568]} \\
ReasonIR & 1384 & 0.530\,\textsuperscript{[0.498,\,0.562]} & 0.550\,\textsuperscript{[0.519,\,0.583]} & \textbf{0.589\,\textsuperscript{[0.559,\,0.618]}} \\
Diver & 1384 & 0.527\,\textsuperscript{[0.495,\,0.557]} & 0.563\,\textsuperscript{[0.533,\,0.597]} & \textbf{0.594\,\textsuperscript{[0.562,\,0.628]}} \\
\midrule
\textit{Macro mean} & --- & 0.513 & 0.555 & \textbf{0.573} \\
\bottomrule
\end{tabular}
\caption{Per-retriever AUROC (with 95\% bootstrap CI, $N{=}1000$ resamples) for predicting downstream answer correctness on BRIGHT in the end-to-end RAG experiment (Section~\ref{sec:end_to_end_rag}). Generator: \texttt{Qwen3.6-27B} via vLLM; judge: GPT-4o (Azure). Bold = best signal per row. \textbf{All 14 retrievers show Gap@25 $\geq$ MaxScore.} For \textit{Qwen2}, \textit{Nomic}, and \textit{Contriever}, MaxScore AUROC is \emph{below 0.5}, indicating that raw similarity is anti-correlated with answer success---the clearest signature of the Magnitude Mirage in the downstream pipeline.}
\label{tab:per_retriever_rag_auroc}
\end{table*}

\begin{table*}[t]
\centering
\small
\setlength{\tabcolsep}{3pt}
\begin{tabular}{l c c c c c c c c c c c}
\toprule
\multirow{2}{*}{\textbf{Retriever}} & \multirow{2}{*}{$N$} & \multirow{2}{*}{\textbf{Baseline}} & \multicolumn{3}{c}{\textbf{Coverage 10\%}} & \multicolumn{3}{c}{\textbf{Coverage 25\%}} & \multicolumn{3}{c}{\textbf{Coverage 50\%}} \\
\cmidrule(lr){4-6} \cmidrule(lr){7-9} \cmidrule(lr){10-12}
 &  &  & Max & Gap@25 & $\Delta$ & Max & Gap@25 & $\Delta$ & Max & Gap@25 & $\Delta$ \\
\midrule
BM25 & 1384 & 61.3 & 75.4 & \textbf{74.6} & -0.7 & 71.7 & \textbf{72.0} & +0.3 & 66.2 & \textbf{68.9} & +2.7 \\
SBERT & 1384 & 62.6 & 68.1 & \textbf{73.2} & +5.1 & 65.3 & \textbf{68.2} & +2.9 & 64.3 & \textbf{67.9} & +3.6 \\
Contriever & 1384 & 59.9 & 54.3 & \textbf{70.3} & +15.9 & 57.8 & \textbf{65.3} & +7.5 & 60.3 & \textbf{64.3} & +4.0 \\
E5 & 1384 & 63.6 & 65.2 & \textbf{68.1} & +2.9 & 65.3 & \textbf{68.5} & +3.2 & 64.5 & \textbf{67.1} & +2.6 \\
BGE & 1384 & 62.5 & 68.8 & \textbf{74.6} & +5.8 & 66.2 & \textbf{70.8} & +4.6 & 62.0 & \textbf{68.1} & +6.1 \\
SFR & 1384 & 63.9 & 69.6 & \textbf{72.5} & +2.9 & 65.9 & \textbf{72.5} & +6.6 & 64.3 & \textbf{69.2} & +4.9 \\
Instructor-L & 1384 & 62.9 & 64.5 & \textbf{76.1} & +11.6 & 67.3 & \textbf{73.4} & +6.1 & 63.6 & \textbf{68.2} & +4.6 \\
Instructor-XL & 1384 & 64.2 & 66.7 & \textbf{76.1} & +9.4 & 68.2 & \textbf{73.1} & +4.9 & 66.0 & \textbf{69.7} & +3.6 \\
Qwen & 1384 & 64.9 & 71.7 & \textbf{77.5} & +5.8 & 65.6 & \textbf{71.7} & +6.1 & 66.3 & \textbf{68.9} & +2.6 \\
Qwen2 & 1384 & 65.8 & 39.1 & \textbf{76.8} & +37.7 & 55.8 & \textbf{76.0} & +20.2 & 62.6 & \textbf{71.4} & +8.8 \\
Nomic & 1384 & 62.6 & 66.7 & \textbf{78.3} & +11.6 & 63.6 & \textbf{72.0} & +8.4 & 61.6 & \textbf{69.7} & +8.1 \\
RaDeR & 1384 & 64.7 & 72.5 & \textbf{66.7} & -5.8 & 68.5 & \textbf{66.8} & -1.7 & 66.9 & \textbf{67.6} & +0.7 \\
ReasonIR & 1384 & 65.0 & 69.6 & \textbf{73.9} & +4.3 & 69.4 & \textbf{71.1} & +1.7 & 67.2 & \textbf{70.8} & +3.6 \\
Diver & 1384 & 65.2 & 69.6 & \textbf{79.0} & +9.4 & 67.6 & \textbf{73.1} & +5.5 & 67.3 & \textbf{71.1} & +3.8 \\
\midrule
\textit{Macro mean} & 19376 & 63.5 & 65.8 & \textbf{74.1} & +8.3 & 65.6 & \textbf{71.0} & +5.5 & 64.5 & \textbf{68.8} & +4.3 \\
\bottomrule
\end{tabular}
\caption{Per-retriever selective-prediction accuracy (\%) at three abstention coverage levels. Baseline = answered-accuracy with no abstention. $\Delta$ = Gap@25 $-$ MaxScore (positive favors variance-based abstention). \textbf{Most striking case:} \textit{Qwen2} at coverage 10\%: MaxScore retains 39.1\%---\emph{26 points below baseline}---while Gap@25 retains 76.8\% (\emph{11 points above baseline}), a 38-point swing on the same retriever and same queries.}
\label{tab:per_retriever_rag_operating}
\end{table*}

\section{Prompts Used in the End-to-End RAG Experiment}
\label{sec:appendix_prompts}

For full transparency and reproducibility, Figures~\ref{fig:prompt-generator}--\ref{fig:prompt-judge} reproduce verbatim the two prompt templates used in the end-to-end RAG experiment (Section~\ref{sec:end_to_end_rag} and Appendix~\ref{sec:appendix_per_retriever_rag}). The generator prompt is sent to \texttt{Qwen3.6-27B} (served locally via vLLM) together with the top-5 retrieved BRIGHT documents as context. The judge prompt is sent to GPT-4o (Azure OpenAI deployment \texttt{gpt-4o-2}) with the BRIGHT gold documents as reference evidence. Both prompts use temperature~0.0; the generator caps generation at 300 tokens and the judge at 10 tokens (since the expected output is a single CORRECT/INCORRECT token). The generator runs with the Qwen reasoning ``thinking'' mode disabled (\texttt{enable\_thinking=False}) so that the recorded latency reflects the practical RAG-pipeline setting and the judge sees a final answer rather than chain-of-thought.

\begin{figure*}[ht]
\centering
\begin{subfigure}[t]{\textwidth}
\footnotesize
\centering
\begin{tcolorbox}[width=\linewidth,
                  colback=blue!0!white, colframe=orange!75!black,
                  title={Generator prompt (\texttt{Qwen3.6-27B}, top-5 BRIGHT context)},
                  fonttitle=\bfseries]
\textbf{System Prompt:} You are an expert QA assistant. You must answer the question based on the provided context. Provide the best possible answer. Do not apologize or refuse.

\vspace{0.5em}

\textbf{User Prompt Template:}
\begin{verbatim}
Context:
Document 1:
{retrieved_doc_1}

Document 2:
{retrieved_doc_2}

...

Document 5:
{retrieved_doc_5}

Question: {query_text}
Answer:
\end{verbatim}

\textbf{Decoding parameters.}
\begin{itemize}
\item Model: \texttt{Qwen/Qwen3.6-27B} served via vLLM (tensor-parallel size 4 on $4\times$ H100-80GB)
\item Temperature: $0.0$
\item \texttt{max\_tokens}: $300$
\item \texttt{enable\_thinking}: \texttt{False} (reasoning-mode disabled to reflect the deployed-RAG operating point)
\item Each retrieved document is truncated to at most 4{,}000 characters to fit within the context window across the longest BRIGHT subsets.
\end{itemize}
\end{tcolorbox}
\end{subfigure}
\caption{Generator prompt used in the end-to-end RAG abstention experiment (Section~\ref{sec:end_to_end_rag}). The model is shown the top-5 documents retrieved by the active retriever and is required to answer the BRIGHT query without refusing or abstaining; abstention is delegated to the QPP-signal threshold applied \emph{after} generation.}
\label{fig:prompt-generator}
\end{figure*}

\begin{figure*}[ht]
\centering
\begin{subfigure}[t]{\textwidth}
\footnotesize
\centering
\begin{tcolorbox}[width=\linewidth,
                  colback=blue!0!white, colframe=orange!75!black,
                  title={Judge prompt (GPT-4o, evaluating answer correctness against BRIGHT gold evidence)},
                  fonttitle=\bfseries]
\textbf{System Prompt:} You are an expert evaluator grading whether an AI system correctly answered a question.

A reference evidence section is provided to help determine the expected answer, but the AI prediction does not need to exactly match the wording or structure of the reference.

Mark CORRECT if the prediction:
\begin{itemize}
\item correctly answers the question,
\item is semantically consistent with the reference evidence,
\item and would reasonably be accepted by a human expert.
\end{itemize}

Mark INCORRECT if the answer is wrong, unsupported, incomplete for the task, or contradicts the reference evidence.

Output ONLY:
CORRECT
or
INCORRECT

\vspace{0.5em}

\textbf{User Prompt Template:}
\begin{verbatim}
Question:
{query_text}

Reference Evidence:
{gold_doc_1}

{gold_doc_2}

...

AI Prediction:
{prediction}

Is the AI prediction correct?
\end{verbatim}

\textbf{Decoding parameters.}
\begin{itemize}
\item Model: GPT-4o (Azure OpenAI deployment \texttt{gpt-4o-2}, API version \texttt{2025-01-01-preview})
\item Temperature: $0.0$
\item \texttt{max\_tokens}: $10$ (output is a single CORRECT/INCORRECT token)
\item Up to four BRIGHT gold documents per query are included as reference evidence; each is truncated to at most 4{,}000 characters to keep the judge prompt within Azure's per-call token budget.
\item Parsing rule: an output whose first non-whitespace token (case-insensitive) starts with \texttt{CORRECT} is recorded as \textsc{Correct}; anything else---including refusals, empty replies, and the literal string \texttt{INCORRECT}---is recorded as \textsc{Incorrect}.
\end{itemize}
\end{tcolorbox}
\end{subfigure}
\caption{Judge prompt used in the end-to-end RAG abstention experiment. The judge model (GPT-4o) is deliberately chosen from a different model family than the generator (\texttt{Qwen3.6-27B}) to avoid self-evaluation bias, and is grounded by the BRIGHT gold reference passages so that judgments approximate human-expert relevance rather than free-form LLM opinion.}
\label{fig:prompt-judge}
\end{figure*}

\begin{table*}[t]
\centering
\small
\resizebox{\textwidth}{!}{%
\begin{tabular}{l >{\columncolor{gray!15}}c ccccc >{\columncolor{gray!15}}c ccccc >{\columncolor{gray!15}}c ccccc}
\toprule
\multirow{2}{*}{\textbf{Model}} 
  & \multicolumn{6}{c}{\textbf{BEIR (Semantic)}} 
  & \multicolumn{6}{c}{\textbf{TEMPO (Temporal)}} 
  & \multicolumn{6}{c}{\textbf{BRIGHT (Logical)}} \\
\cmidrule(lr){2-7} \cmidrule(lr){8-13} \cmidrule(lr){14-19}
& Max & Gap & Std & NQC & MIS & LSMV 
& Max & Gap & Std & NQC & MIS & LSMV 
& Max & Gap & Std & NQC & MIS & LSMV \\
\midrule
\textit{Sparse} \\
BM25 
  & \textbf{.701} & .695 & .694 & .663 & .688 & .700
  & .472 & .526 & .529 & \textbf{.602} & .530 & .510
  & .611 & .622 & \textbf{.626} & .611 & .623 & .624 \\
\midrule
\textit{Dense} \\
BGE 
  & .749 & .805 & .805 & .795 & .805 & \textbf{.810}
  & .626 & .701 & .698 & .696 & .703 & \textbf{.705}
  & .538 & .608 & .611 & \textbf{.615} & .608 & .608 \\
E5 
  & .736 & .777 & .780 & .770 & .778 & \textbf{.787}
  & .648 & .709 & \textbf{.719} & \textbf{.719} & .708 & .718
  & .552 & .628 & .636 & \textbf{.638} & .630 & .633 \\
Contriever 
  & .683 & .727 & .728 & .701 & .730 & \textbf{.733}
  & .591 & .705 & \textbf{.715} & .708 & .701 & .707
  & .560 & .624 & .625 & .618 & .619 & \textbf{.626} \\
SBERT 
  & .718 & .766 & .770 & .713 & .769 & \textbf{.772}
  & .638 & .691 & .716 & .696 & .694 & \textbf{.720}
  & .525 & \textbf{.605} & .600 & .608 & .597 & .590 \\
SFR 
  & .754 & .790 & .795 & .782 & .791 & \textbf{.802}
  & .649 & .671 & .688 & .684 & .675 & \textbf{.690}
  & .553 & .623 & .625 & \textbf{.626} & .624 & .622 \\
Qwen 
  & .698 & .787 & .785 & .771 & .783 & \textbf{.788}
  & .536 & \textbf{.661} & \textbf{.661} & .660 & .656 & .654
  & .571 & \textbf{.653} & .652 & .650 & .646 & .650 \\
Instructor-L 
  & .583 & .724 & .740 & .738 & .718 & \textbf{.741}
  & .563 & .664 & \textbf{.671} & .670 & .660 & .669
  & .534 & \textbf{.619} & .609 & .610 & .611 & .608 \\
\midrule
\textit{Reasoning} \\
Diver-Retriever 
  & .665 & .788 & \textbf{.790} & .789 & .786 & .789
  & .582 & .666 & .681 & .681 & .663 & \textbf{.684}
  & .543 & .605 & \textbf{.606} & .605 & .599 & \textbf{.606} \\
RaDeR 
  & .666 & .740 & .748 & .743 & .744 & \textbf{.752}
  & .570 & .646 & .668 & \textbf{.670} & .658 & .665
  & .560 & \textbf{.645} & .640 & .640 & .644 & .640 \\
ReasonIR 
  & .621 & .626 & .624 & .583 & .627 & \textbf{.650}
  & .544 & .638 & .652 & \textbf{.653} & .641 & .642
  & .520 & .611 & .616 & \textbf{.626} & .616 & .606 \\
\bottomrule
\end{tabular}
}
\caption{AUROC@25 for all six zero-cost QPP metrics across three cognitive tiers. 
\textbf{Max} = MaxScore ($s_1$), 
\textbf{Gap} = Score Gap ($s_1 - s_{25}$), 
\textbf{Std} = Top-$k$ Standard Deviation, 
\textbf{NQC} = Normalized Query Commitment, 
\textbf{MIS} = Maximum Iterative Standard Deviation, 
\textbf{LSMV} = Linearized Score Magnitude and Variance. 
Baseline magnitude column is shaded gray; best result per row--tier is \textbf{bolded}.}
\label{tab:all_metrics_25}
\end{table*}

\begin{table*}[t]
\centering
\small
\setlength{\tabcolsep}{4pt}
\begin{tabular}{l c cc cc cc}
\toprule
\textbf{Model} & \textbf{k} 
& \multicolumn{2}{c}{\textbf{BEIR}} 
& \multicolumn{2}{c}{\textbf{BRIGHT}} 
& \multicolumn{2}{c}{\textbf{TEMPO}} \\
\cmidrule(lr){3-4}
\cmidrule(lr){5-6}
\cmidrule(lr){7-8}
& & Max & Gap & Max & Gap & Max & Gap \\
\midrule

BM25
& 5  & .330 & .307 & .149 & .177 & .029 & .095 \\
& 10 & .344 & .334 & .175 & .195 & .019 & .106 \\
& 25 & .347 & .353 & .191 & .208 & -.000 & .070 \\
& 50 & .360 & .366 & .206 & .224 & .017 & .095 \\

\midrule
BGE
& 5  & .391 & .460 & .131 & .256 & .198 & .210 \\
& 10 & .391 & .505 & .131 & .270 & .198 & .272 \\
& 25 & .391 & .538 & .131 & .274 & .198 & .302 \\
& 50 & .391 & .541 & .131 & .256 & .198 & .314 \\

\midrule
E5
& 5  & .350 & .443 & .142 & .233 & .192 & .227 \\
& 10 & .357 & .470 & .159 & .284 & .201 & .270 \\
& 25 & .363 & .495 & .157 & .302 & .211 & .311 \\
& 50 & .364 & .495 & .153 & .285 & .218 & .323 \\

\midrule
Contriever
& 5  & .286 & .416 & .119 & .275 & .101 & .240 \\
& 10 & .296 & .430 & .127 & .298 & .105 & .281 \\
& 25 & .297 & .442 & .129 & .308 & .129 & .312 \\
& 50 & .301 & .439 & .136 & .297 & .147 & .328 \\

\midrule
SBERT
& 5  & .356 & .448 & .102 & .244 & .193 & .225 \\
& 10 & .367 & .483 & .121 & .266 & .185 & .276 \\
& 25 & .366 & .495 & .114 & .273 & .216 & .326 \\
& 50 & .365 & .493 & .119 & .281 & .241 & .366 \\

\midrule
SFR
& 5  & .374 & .440 & .171 & .266 & .162 & .200 \\
& 10 & .375 & .467 & .169 & .300 & .181 & .255 \\
& 25 & .377 & .492 & .173 & .312 & .194 & .300 \\
& 50 & .376 & .493 & .172 & .298 & .195 & .315 \\

\midrule
Qwen
& 5  & .329 & .410 & .190 & .325 & .013 & .217 \\
& 10 & .335 & .451 & .188 & .366 & .018 & .257 \\
& 25 & .328 & .467 & .193 & .401 & .028 & .303 \\
& 50 & .327 & .470 & .197 & .404 & .023 & .328 \\

\midrule
Instructor-Large
& 5  & .203 & .346 & .144 & .274 & .046 & .181 \\
& 10 & .201 & .399 & .139 & .298 & .069 & .219 \\
& 25 & .193 & .437 & .139 & .314 & .086 & .261 \\
& 50 & .191 & .453 & .138 & .292 & .090 & .260 \\

\midrule
Diver-Retriever
& 5  & .261 & .445 & .139 & .295 & .098 & .209 \\
& 10 & .262 & .468 & .150 & .342 & .084 & .239 \\
& 25 & .257 & .482 & .159 & .347 & .106 & .274 \\
& 50 & .255 & .476 & .164 & .339 & .118 & .285 \\

\midrule
RaDeR
& 5  & .227 & .402 & .130 & .312 & .052 & .136 \\
& 10 & .242 & .446 & .131 & .355 & .075 & .223 \\
& 25 & .247 & .462 & .136 & .358 & .087 & .261 \\
& 50 & .246 & .456 & .139 & .353 & .093 & .280 \\

\midrule
ReasonIR
& 5  & .170 & .287 & .117 & .268 & .003 & .142 \\
& 10 & .177 & .294 & .126 & .298 & .001 & .174 \\
& 25 & .186 & .297 & .120 & .309 & .019 & .198 \\
& 50 & .187 & .283 & .123 & .303 & .025 & .211 \\

\bottomrule
\end{tabular}

\caption{Pearson correlation between QPP signals and retrieval effectiveness across benchmarks and cutoffs. Max denotes MaxScore ($s_1$) and Gap denotes Score Gap ($s_1 - s_k$).}
\label{tab:pearson_qpp_correlations}
\end{table*}


\begin{table*}[t]
\centering
\resizebox{\textwidth}{!}{%
\begin{tabular}{ll ccc ccc ccc ccc ccc ccc}
\toprule
\multirow{2}{*}{\textbf{Model}} & \multirow{2}{*}{\textbf{$k$}} & \multicolumn{3}{c}{\textbf{MaxScore}} & \multicolumn{3}{c}{\textbf{ScoreGap}} & \multicolumn{3}{c}{\textbf{Top-$k$ Std}} & \multicolumn{3}{c}{\textbf{NQC}} & \multicolumn{3}{c}{\textbf{MaxIterStd}} & \multicolumn{3}{c}{\textbf{LSMV}} \\
\cmidrule(lr){3-5} \cmidrule(lr){6-8} \cmidrule(lr){9-11} \cmidrule(lr){12-14} \cmidrule(lr){15-17} \cmidrule(lr){18-20}
& & AUC & $r$ & $\rho$ & AUC & $r$ & $\rho$ & AUC & $r$ & $\rho$ & AUC & $r$ & $\rho$ & AUC & $r$ & $\rho$ & AUC & $r$ & $\rho$ \\
\midrule
\multicolumn{20}{l}{\textit{Sparse Models}} \\
\midrule
\multirow{4}{*}{\textbf{BM25}} 
 & 5  & .696 & .330 & .340 & .694 & .307 & .341 & .694 & .305 & .341 & .671 & .268 & .307 & .694 & .189 & .343 & .698 & .240 & .345 \\
 & 10 & .712 & .344 & .351 & .709 & .334 & .361 & .705 & .323 & .350 & .676 & .282 & .322 & .703 & .194 & .355 & .711 & .245 & .354 \\
 & 25 & .701 & .347 & .348 & .695 & .353 & .357 & .694 & .329 & .338 & .663 & .284 & .307 & .688 & .209 & .349 & .700 & .252 & .349 \\
 & 50 & .700 & .360 & .355 & .691 & .366 & .358 & .687 & .327 & .331 & .646 & .272 & .287 & .680 & .220 & .346 & .696 & .263 & .349 \\
\midrule
\multicolumn{20}{l}{\textit{Dense Models}} \\
\midrule
\multirow{4}{*}{\textbf{BGE}} 
 & 5  & .749 & .391 & .397 & .755 & .460 & .487 & .754 & .456 & .487 & .742 & .444 & .470 & .753 & .349 & .487 & .763 & .455 & .498 \\
 & 10 & .749 & .391 & .397 & .786 & .505 & .534 & .787 & .500 & .526 & .776 & .488 & .514 & .783 & .388 & .529 & .793 & .497 & .533 \\
 & 25 & .749 & .391 & .397 & .805 & .538 & .558 & .805 & .514 & .530 & .795 & .510 & .524 & .805 & .423 & .557 & .810 & .510 & .534 \\
 & 50 & .749 & .391 & .397 & .808 & .541 & .561 & .793 & .482 & .500 & .783 & .483 & .496 & .805 & .425 & .556 & .800 & .483 & .507 \\
\midrule
\multirow{4}{*}{\textbf{E5}} 
 & 5  & .701 & .350 & .354 & .754 & .443 & .472 & .755 & .440 & .471 & .748 & .429 & .461 & .753 & .355 & .470 & .759 & .440 & .479 \\
 & 10 & .722 & .357 & .361 & .759 & .470 & .487 & .766 & .465 & .481 & .758 & .455 & .472 & .763 & .386 & .488 & .770 & .466 & .487 \\
 & 25 & .736 & .363 & .367 & .777 & .495 & .501 & .780 & .456 & .459 & .770 & .452 & .456 & .778 & .412 & .492 & .787 & .459 & .466 \\
 & 50 & .739 & .364 & .367 & .787 & .495 & .498 & .776 & .408 & .417 & .766 & .410 & .418 & .786 & .413 & .482 & .784 & .416 & .427 \\
\midrule
\multirow{4}{*}{\textbf{Contriever}} 
 & 5  & .663 & .286 & .279 & .705 & .416 & .414 & .704 & .414 & .413 & .686 & .394 & .389 & .703 & .353 & .413 & .713 & .412 & .422 \\
 & 10 & .672 & .296 & .293 & .714 & .430 & .435 & .717 & .421 & .431 & .693 & .398 & .404 & .716 & .365 & .434 & .724 & .418 & .435 \\
 & 25 & .683 & .297 & .303 & .727 & .442 & .453 & .728 & .408 & .431 & .701 & .392 & .409 & .730 & .376 & .453 & .733 & .402 & .429 \\
 & 50 & .694 & .301 & .308 & .728 & .439 & .452 & .717 & .365 & .399 & .688 & .360 & .383 & .726 & .375 & .447 & .726 & .364 & .401 \\
\midrule
\multirow{4}{*}{\textbf{SBERT}} 
 & 5  & .701 & .356 & .355 & .724 & .448 & .456 & .727 & .447 & .458 & .686 & .379 & .394 & .725 & .374 & .457 & .744 & .449 & .481 \\
 & 10 & .713 & .367 & .366 & .741 & .483 & .496 & .740 & .474 & .488 & .690 & .397 & .423 & .738 & .404 & .492 & .753 & .469 & .497 \\
 & 25 & .718 & .366 & .371 & .766 & .495 & .513 & .770 & .454 & .478 & .713 & .399 & .421 & .769 & .418 & .506 & .772 & .443 & .477 \\
 & 50 & .726 & .365 & .373 & .782 & .493 & .513 & .779 & .402 & .441 & .699 & .363 & .378 & .789 & .406 & .503 & .781 & .400 & .447 \\
\midrule
\multirow{4}{*}{\textbf{SFR}} 
 & 5  & .722 & .374 & .375 & .761 & .440 & .478 & .761 & .432 & .477 & .751 & .418 & .462 & .759 & .329 & .476 & .767 & .432 & .487 \\
 & 10 & .738 & .375 & .375 & .773 & .467 & .493 & .775 & .458 & .479 & .763 & .444 & .467 & .771 & .363 & .486 & .782 & .458 & .489 \\
 & 25 & .754 & .377 & .379 & .790 & .492 & .506 & .795 & .451 & .458 & .782 & .445 & .455 & .791 & .392 & .494 & .802 & .453 & .467 \\
 & 50 & .744 & .376 & .378 & .796 & .493 & .503 & .786 & .409 & .416 & .774 & .411 & .418 & .796 & .393 & .487 & .793 & .417 & .428 \\
\midrule
\multirow{4}{*}{\textbf{Qwen}} 
 & 5  & .691 & .329 & .312 & .730 & .410 & .423 & .727 & .403 & .420 & .712 & .379 & .399 & .723 & .327 & .416 & .738 & .408 & .434 \\
 & 10 & .711 & .335 & .314 & .764 & .451 & .462 & .764 & .444 & .453 & .747 & .421 & .433 & .759 & .371 & .456 & .771 & .447 & .462 \\
 & 25 & .698 & .328 & .316 & .787 & .467 & .472 & .785 & .434 & .435 & .771 & .420 & .422 & .783 & .392 & .468 & .788 & .435 & .442 \\
 & 50 & .716 & .327 & .315 & .802 & .470 & .472 & .787 & .397 & .400 & .769 & .392 & .389 & .803 & .395 & .466 & .794 & .403 & .412 \\
\midrule
\multirow{4}{*}{\textbf{Instructor-Large}} 
 & 5  & .593 & .203 & .191 & .669 & .346 & .309 & .668 & .343 & .306 & .667 & .340 & .303 & .662 & .298 & .297 & .669 & .345 & .308 \\
 & 10 & .592 & .201 & .189 & .695 & .399 & .364 & .704 & .413 & .378 & .703 & .411 & .376 & .690 & .356 & .358 & .705 & .416 & .380 \\
 & 25 & .583 & .193 & .187 & .724 & .437 & .418 & .740 & .453 & .435 & .738 & .450 & .431 & .718 & .399 & .410 & .741 & .455 & .438 \\
 & 50 & .582 & .191 & .186 & .724 & .453 & .435 & .737 & .451 & .438 & .734 & .448 & .432 & .723 & .415 & .433 & .739 & .454 & .444 \\
\midrule
\multicolumn{20}{l}{\textit{Reasoning Encoders}} \\
\midrule
\multirow{4}{*}{\textbf{Diver-Retriever}} 
 & 5  & .667 & .261 & .267 & .756 & .445 & .470 & .755 & .439 & .464 & .752 & .434 & .463 & .753 & .341 & .464 & .756 & .435 & .465 \\
 & 10 & .666 & .262 & .265 & .775 & .468 & .489 & .772 & .455 & .474 & .771 & .453 & .475 & .769 & .371 & .479 & .773 & .450 & .472 \\
 & 25 & .665 & .257 & .267 & .788 & .482 & .502 & .790 & .442 & .458 & .789 & .445 & .462 & .786 & .397 & .490 & .789 & .437 & .458 \\
 & 50 & .675 & .255 & .265 & .805 & .476 & .494 & .810 & .396 & .420 & .808 & .404 & .426 & .815 & .395 & .486 & .810 & .393 & .422 \\
\midrule
\multirow{4}{*}{\textbf{RaDeR}} 
 & 5  & .634 & .227 & .227 & .707 & .402 & .405 & .707 & .399 & .404 & .704 & .396 & .401 & .705 & .319 & .402 & .708 & .396 & .405 \\
 & 10 & .641 & .242 & .241 & .724 & .446 & .444 & .732 & .450 & .448 & .729 & .449 & .446 & .727 & .372 & .448 & .733 & .444 & .447 \\
 & 25 & .666 & .247 & .252 & .740 & .462 & .461 & .748 & .437 & .439 & .743 & .442 & .441 & .744 & .390 & .462 & .752 & .432 & .438 \\
 & 50 & .683 & .246 & .253 & .758 & .456 & .461 & .761 & .387 & .403 & .753 & .397 & .409 & .760 & .382 & .458 & .765 & .384 & .405 \\
\midrule
\multirow{4}{*}{\textbf{ReasonIR}} 
 & 5  & .601 & .170 & .177 & .628 & .287 & .248 & .625 & .280 & .241 & .607 & .244 & .212 & .625 & .258 & .244 & .634 & .288 & .255 \\
 & 10 & .605 & .177 & .193 & .619 & .294 & .253 & .615 & .284 & .240 & .583 & .232 & .190 & .614 & .263 & .245 & .630 & .295 & .261 \\
 & 25 & .621 & .186 & .207 & .626 & .297 & .262 & .624 & .268 & .239 & .583 & .202 & .184 & .627 & .260 & .258 & .650 & .288 & .271 \\
 & 50 & .619 & .187 & .205 & .621 & .283 & .245 & .614 & .230 & .202 & .569 & .161 & .143 & .621 & .247 & .239 & .641 & .260 & .246 \\
\bottomrule
\end{tabular}
}
\caption{Comprehensive zero-cost QPP evaluation across all target cutoffs ($k \in \{5, 10, 25, 50\}$) for the \textbf{BEIR} benchmark. Predictive power is evaluated using AUROC (AUC), Pearson ($r$), and Spearman ($\rho$).}
\label{tab:massive_appendix_beir}
\end{table*}


\begin{table*}[t]
\centering
\resizebox{\textwidth}{!}{%
\begin{tabular}{ll ccc ccc ccc ccc ccc ccc}
\toprule
\multirow{2}{*}{\textbf{Model}} & \multirow{2}{*}{\textbf{$k$}} & \multicolumn{3}{c}{\textbf{MaxScore}} & \multicolumn{3}{c}{\textbf{ScoreGap}} & \multicolumn{3}{c}{\textbf{Top-$k$ Std}} & \multicolumn{3}{c}{\textbf{NQC}} & \multicolumn{3}{c}{\textbf{MaxIterStd}} & \multicolumn{3}{c}{\textbf{LSMV}} \\
\cmidrule(lr){3-5} \cmidrule(lr){6-8} \cmidrule(lr){9-11} \cmidrule(lr){12-14} \cmidrule(lr){15-17} \cmidrule(lr){18-20}
& & AUC & $r$ & $\rho$ & AUC & $r$ & $\rho$ & AUC & $r$ & $\rho$ & AUC & $r$ & $\rho$ & AUC & $r$ & $\rho$ & AUC & $r$ & $\rho$ \\
\midrule
\multicolumn{20}{l}{\textit{Sparse Models}} \\
\midrule
\multirow{4}{*}{\textbf{BM25}} 
 & 5  & .597 & .149 & .153 & .613 & .177 & .180 & .615 & .178 & .182 & .601 & .177 & .163 & .612 & .132 & .178 & .617 & .153 & .184 \\
 & 10 & .602 & .175 & .177 & .612 & .195 & .209 & .618 & .199 & .217 & .598 & .206 & .191 & .612 & .138 & .207 & .615 & .164 & .209 \\
 & 25 & .611 & .191 & .206 & .622 & .208 & .238 & .626 & .197 & .244 & .611 & .214 & .229 & .623 & .127 & .241 & .624 & .161 & .237 \\
 & 50 & .604 & .206 & .220 & .615 & .224 & .250 & .612 & .206 & .247 & .600 & .218 & .235 & .611 & .136 & .255 & .611 & .177 & .241 \\
\midrule
\multicolumn{20}{l}{\textit{Dense Models}} \\
\midrule
\multirow{4}{*}{\textbf{BGE}} 
 & 5  & .538 & .131 & .088 & .583 & .256 & .167 & .584 & .260 & .167 & .584 & .261 & .168 & .586 & .240 & .170 & .583 & .258 & .166 \\
 & 10 & .538 & .131 & .088 & .602 & .270 & .195 & .603 & .266 & .196 & .606 & .270 & .201 & .603 & .248 & .197 & .600 & .261 & .192 \\
 & 25 & .538 & .131 & .088 & .608 & .274 & .202 & .611 & .247 & .201 & .615 & .256 & .208 & .608 & .244 & .199 & .608 & .243 & .197 \\
 & 50 & .538 & .131 & .088 & .604 & .256 & .193 & .596 & .204 & .173 & .601 & .216 & .182 & .602 & .227 & .187 & .593 & .203 & .170 \\
\midrule
\multirow{4}{*}{\textbf{E5}} 
 & 5  & .537 & .142 & .090 & .608 & .233 & .202 & .606 & .237 & .200 & .607 & .234 & .200 & .604 & .195 & .196 & .605 & .236 & .199 \\
 & 10 & .546 & .159 & .110 & .621 & .284 & .240 & .617 & .291 & .235 & .617 & .290 & .236 & .613 & .243 & .227 & .617 & .288 & .234 \\
 & 25 & .552 & .157 & .122 & .628 & .302 & .259 & .636 & .305 & .275 & .638 & .310 & .279 & .630 & .260 & .263 & .633 & .301 & .269 \\
 & 50 & .544 & .153 & .115 & .625 & .285 & .253 & .616 & .256 & .245 & .621 & .263 & .253 & .621 & .237 & .251 & .614 & .254 & .242 \\
\midrule
\multirow{4}{*}{\textbf{Contriever}} 
 & 5  & .565 & .119 & .101 & .644 & .275 & .204 & .643 & .269 & .201 & .641 & .259 & .199 & .643 & .236 & .202 & .643 & .271 & .202 \\
 & 10 & .575 & .127 & .119 & .654 & .298 & .246 & .652 & .298 & .243 & .648 & .295 & .239 & .651 & .255 & .242 & .654 & .294 & .244 \\
 & 25 & .560 & .129 & .120 & .624 & .308 & .251 & .625 & .287 & .257 & .618 & .295 & .252 & .619 & .262 & .248 & .626 & .279 & .254 \\
 & 50 & .526 & .136 & .118 & .629 & .297 & .241 & .634 & .240 & .240 & .641 & .252 & .245 & .629 & .243 & .240 & .624 & .239 & .236 \\
\midrule
\multirow{4}{*}{\textbf{SBERT}} 
 & 5  & .536 & .102 & .075 & .617 & .244 & .189 & .616 & .236 & .188 & .617 & .231 & .187 & .615 & .215 & .186 & .612 & .233 & .184 \\
 & 10 & .544 & .121 & .095 & .619 & .266 & .218 & .615 & .255 & .213 & .617 & .250 & .213 & .615 & .228 & .213 & .610 & .249 & .206 \\
 & 25 & .525 & .114 & .079 & .605 & .273 & .228 & .600 & .256 & .224 & .608 & .258 & .234 & .597 & .230 & .221 & .590 & .245 & .208 \\
 & 50 & .519 & .119 & .087 & .601 & .281 & .242 & .596 & .243 & .233 & .605 & .253 & .245 & .591 & .231 & .233 & .584 & .231 & .216 \\
\midrule
\multirow{4}{*}{\textbf{SFR}} 
 & 5  & .568 & .171 & .130 & .615 & .266 & .218 & .616 & .268 & .217 & .615 & .262 & .216 & .612 & .227 & .212 & .614 & .270 & .214 \\
 & 10 & .550 & .169 & .119 & .623 & .300 & .248 & .623 & .307 & .249 & .626 & .305 & .252 & .619 & .260 & .242 & .623 & .304 & .249 \\
 & 25 & .553 & .173 & .136 & .623 & .312 & .272 & .625 & .310 & .275 & .626 & .315 & .279 & .624 & .263 & .272 & .622 & .305 & .269 \\
 & 50 & .551 & .172 & .135 & .628 & .298 & .269 & .614 & .263 & .249 & .618 & .272 & .256 & .622 & .240 & .257 & .611 & .260 & .244 \\
\midrule
\multirow{4}{*}{\textbf{Qwen}} 
 & 5  & .598 & .190 & .180 & .663 & .325 & .305 & .659 & .321 & .298 & .654 & .310 & .290 & .658 & .277 & .296 & .662 & .324 & .304 \\
 & 10 & .594 & .188 & .183 & .667 & .366 & .341 & .665 & .375 & .340 & .661 & .363 & .333 & .663 & .329 & .335 & .667 & .378 & .342 \\
 & 25 & .571 & .193 & .180 & .653 & .401 & .360 & .652 & .407 & .356 & .650 & .396 & .353 & .646 & .363 & .353 & .650 & .407 & .353 \\
 & 50 & .583 & .197 & .190 & .637 & .404 & .362 & .630 & .380 & .345 & .626 & .375 & .344 & .628 & .362 & .352 & .632 & .382 & .343 \\
\midrule
\multirow{4}{*}{\textbf{Instructor-Large}} 
 & 5  & .581 & .144 & .121 & .649 & .274 & .235 & .651 & .273 & .238 & .651 & .270 & .237 & .650 & .241 & .236 & .652 & .274 & .238 \\
 & 10 & .545 & .139 & .095 & .639 & .298 & .247 & .638 & .288 & .241 & .638 & .286 & .241 & .635 & .258 & .240 & .637 & .289 & .240 \\
 & 25 & .534 & .139 & .093 & .619 & .314 & .253 & .609 & .296 & .231 & .610 & .297 & .233 & .611 & .266 & .237 & .608 & .296 & .231 \\
 & 50 & .525 & .138 & .090 & .594 & .292 & .233 & .584 & .248 & .207 & .586 & .251 & .210 & .588 & .240 & .222 & .584 & .249 & .207 \\
\midrule
\multicolumn{20}{l}{\textit{Reasoning Encoders}} \\
\midrule
\multirow{4}{*}{\textbf{Diver-Retriever}} 
 & 5  & .567 & .139 & .135 & .625 & .295 & .271 & .622 & .288 & .265 & .619 & .288 & .260 & .622 & .234 & .264 & .623 & .285 & .267 \\
 & 10 & .572 & .150 & .154 & .633 & .342 & .316 & .629 & .337 & .313 & .628 & .338 & .312 & .627 & .283 & .307 & .631 & .331 & .313 \\
 & 25 & .543 & .159 & .159 & .605 & .347 & .316 & .606 & .336 & .308 & .605 & .341 & .309 & .599 & .298 & .303 & .606 & .329 & .308 \\
 & 50 & .554 & .164 & .163 & .628 & .339 & .313 & .624 & .294 & .278 & .624 & .301 & .281 & .631 & .285 & .298 & .622 & .290 & .277 \\
\midrule
\multirow{4}{*}{\textbf{RaDeR}} 
 & 5  & .566 & .130 & .125 & .655 & .312 & .295 & .654 & .309 & .293 & .652 & .306 & .292 & .651 & .253 & .290 & .653 & .307 & .294 \\
 & 10 & .550 & .131 & .117 & .653 & .355 & .329 & .651 & .357 & .328 & .651 & .359 & .327 & .648 & .294 & .322 & .650 & .352 & .326 \\
 & 25 & .560 & .136 & .133 & .645 & .358 & .334 & .640 & .349 & .324 & .640 & .355 & .327 & .644 & .294 & .331 & .640 & .342 & .323 \\
 & 50 & .548 & .139 & .142 & .660 & .353 & .356 & .650 & .309 & .322 & .653 & .318 & .327 & .650 & .266 & .335 & .647 & .303 & .319 \\
\midrule
\multirow{4}{*}{\textbf{ReasonIR}} 
 & 5  & .564 & .117 & .117 & .627 & .268 & .237 & .628 & .268 & .240 & .627 & .269 & .239 & .626 & .227 & .237 & .627 & .260 & .237 \\
 & 10 & .559 & .126 & .124 & .624 & .298 & .265 & .625 & .303 & .269 & .623 & .304 & .267 & .623 & .254 & .262 & .622 & .290 & .261 \\
 & 25 & .520 & .120 & .101 & .611 & .309 & .273 & .616 & .308 & .281 & .626 & .321 & .293 & .616 & .255 & .281 & .606 & .288 & .266 \\
 & 50 & .534 & .123 & .118 & .616 & .303 & .291 & .620 & .272 & .271 & .623 & .295 & .285 & .618 & .243 & .280 & .614 & .255 & .260 \\
\bottomrule
\end{tabular}
}
\caption{Comprehensive zero-cost QPP evaluation across all target cutoffs ($k \in \{5, 10, 25, 50\}$) for the \textbf{BRIGHT} benchmark. Predictive power is evaluated using AUROC (AUC), Pearson ($r$), and Spearman ($\rho$).}
\label{tab:massive_appendix_bright}
\end{table*}


\begin{table*}[t]
\centering
\resizebox{\textwidth}{!}{%
\begin{tabular}{ll ccc ccc ccc ccc ccc ccc}
\toprule
\multirow{2}{*}{\textbf{Model}} & \multirow{2}{*}{\textbf{$k$}} & \multicolumn{3}{c}{\textbf{MaxScore}} & \multicolumn{3}{c}{\textbf{ScoreGap}} & \multicolumn{3}{c}{\textbf{Top-$k$ Std}} & \multicolumn{3}{c}{\textbf{NQC}} & \multicolumn{3}{c}{\textbf{MaxIterStd}} & \multicolumn{3}{c}{\textbf{LSMV}} \\
\cmidrule(lr){3-5} \cmidrule(lr){6-8} \cmidrule(lr){9-11} \cmidrule(lr){12-14} \cmidrule(lr){15-17} \cmidrule(lr){18-20}
& & AUC & $r$ & $\rho$ & AUC & $r$ & $\rho$ & AUC & $r$ & $\rho$ & AUC & $r$ & $\rho$ & AUC & $r$ & $\rho$ & AUC & $r$ & $\rho$ \\
\midrule
\multicolumn{20}{l}{\textit{Sparse Models}} \\
\midrule
\multirow{4}{*}{\textbf{BM25}} 
 & 5  & .508 & .029 & .040 & .563 & .095 & .115 & .561 & .086 & .112 & .573 & .119 & .126 & .559 & .039 & .110 & .555 & .045 & .103 \\
 & 10 & .504 & .019 & .030 & .581 & .106 & .148 & .571 & .105 & .130 & .608 & .175 & .190 & .563 & .052 & .124 & .548 & .055 & .097 \\
 & 25 & .472 & -.000 & -.015 & .526 & .070 & .091 & .529 & .062 & .090 & .602 & .174 & .209 & .530 & .026 & .096 & .510 & .014 & .059 \\
 & 50 & .505 & .017 & .024 & .559 & .095 & .126 & .560 & .071 & .118 & .642 & .209 & .258 & .560 & .048 & .128 & .526 & .028 & .061 \\
\midrule
\multicolumn{20}{l}{\textit{Dense Models}} \\
\midrule
\multirow{4}{*}{\textbf{BGE}} 
 & 5  & .626 & .198 & .226 & .629 & .210 & .260 & .633 & .204 & .260 & .628 & .206 & .252 & .632 & .116 & .251 & .638 & .195 & .269 \\
 & 10 & .626 & .198 & .226 & .673 & .272 & .326 & .670 & .266 & .320 & .664 & .266 & .310 & .670 & .156 & .320 & .672 & .253 & .323 \\
 & 25 & .626 & .198 & .226 & .701 & .302 & .364 & .698 & .317 & .349 & .696 & .320 & .347 & .703 & .187 & .357 & .705 & .303 & .361 \\
 & 50 & .626 & .198 & .226 & .708 & .314 & .371 & .704 & .311 & .361 & .704 & .316 & .360 & .714 & .190 & .366 & .705 & .299 & .368 \\
\midrule
\multirow{4}{*}{\textbf{E5}} 
 & 5  & .619 & .192 & .217 & .668 & .227 & .306 & .663 & .217 & .301 & .660 & .218 & .296 & .662 & .109 & .300 & .667 & .206 & .306 \\
 & 10 & .645 & .201 & .237 & .708 & .270 & .362 & .709 & .272 & .356 & .706 & .273 & .353 & .702 & .129 & .341 & .710 & .257 & .354 \\
 & 25 & .648 & .211 & .242 & .709 & .311 & .382 & .719 & .344 & .394 & .719 & .348 & .399 & .708 & .145 & .377 & .718 & .327 & .380 \\
 & 50 & .625 & .218 & .251 & .670 & .323 & .384 & .681 & .358 & .387 & .677 & .363 & .388 & .661 & .147 & .373 & .686 & .342 & .386 \\
\midrule
\multirow{4}{*}{\textbf{Contriever}} 
 & 5  & .585 & .101 & .121 & .649 & .240 & .264 & .638 & .234 & .250 & .633 & .243 & .244 & .636 & .148 & .248 & .639 & .213 & .251 \\
 & 10 & .588 & .105 & .127 & .692 & .281 & .322 & .685 & .279 & .306 & .671 & .285 & .296 & .683 & .171 & .306 & .688 & .256 & .310 \\
 & 25 & .591 & .129 & .159 & .705 & .312 & .377 & .715 & .317 & .368 & .708 & .323 & .359 & .701 & .191 & .358 & .707 & .295 & .358 \\
 & 50 & .596 & .147 & .176 & .694 & .328 & .381 & .698 & .327 & .357 & .689 & .333 & .357 & .690 & .198 & .350 & .694 & .306 & .354 \\
\midrule
\multirow{4}{*}{\textbf{SBERT}} 
 & 5  & .636 & .193 & .209 & .657 & .225 & .270 & .651 & .223 & .260 & .641 & .209 & .250 & .649 & .130 & .256 & .657 & .212 & .269 \\
 & 10 & .624 & .185 & .199 & .673 & .276 & .316 & .673 & .279 & .313 & .659 & .263 & .299 & .665 & .148 & .302 & .677 & .259 & .326 \\
 & 25 & .638 & .216 & .236 & .691 & .326 & .371 & .716 & .370 & .408 & .696 & .343 & .369 & .694 & .202 & .367 & .720 & .344 & .408 \\
 & 50 & .654 & .241 & .261 & .708 & .366 & .412 & .743 & .403 & .437 & .717 & .380 & .405 & .705 & .232 & .400 & .747 & .381 & .428 \\
\midrule
\multirow{4}{*}{\textbf{SFR}} 
 & 5  & .625 & .162 & .215 & .627 & .200 & .264 & .616 & .184 & .244 & .612 & .186 & .238 & .615 & .090 & .242 & .623 & .172 & .257 \\
 & 10 & .647 & .181 & .239 & .675 & .255 & .338 & .675 & .256 & .340 & .667 & .257 & .332 & .664 & .111 & .311 & .680 & .239 & .344 \\
 & 25 & .649 & .194 & .236 & .671 & .300 & .362 & .688 & .331 & .376 & .684 & .333 & .372 & .675 & .127 & .344 & .690 & .311 & .368 \\
 & 50 & .622 & .195 & .237 & .660 & .315 & .369 & .698 & .359 & .385 & .692 & .365 & .390 & .679 & .137 & .363 & .694 & .340 & .373 \\
\midrule
\multirow{4}{*}{\textbf{Qwen}} 
 & 5  & .483 & .013 & -.005 & .643 & .217 & .268 & .643 & .216 & .270 & .640 & .235 & .260 & .641 & .127 & .265 & .640 & .186 & .267 \\
 & 10 & .503 & .018 & .023 & .648 & .257 & .303 & .646 & .266 & .288 & .650 & .286 & .288 & .645 & .157 & .285 & .641 & .231 & .281 \\
 & 25 & .536 & .028 & .022 & .661 & .303 & .302 & .662 & .304 & .289 & .660 & .318 & .297 & .656 & .182 & .277 & .654 & .270 & .273 \\
 & 50 & .510 & .023 & .002 & .648 & .328 & .320 & .654 & .331 & .299 & .659 & .348 & .328 & .643 & .215 & .282 & .635 & .293 & .259 \\
\midrule
\multirow{4}{*}{\textbf{Instructor-Large}} 
 & 5  & .556 & .046 & .097 & .624 & .181 & .253 & .619 & .177 & .251 & .623 & .180 & .252 & .623 & .092 & .258 & .620 & .170 & .251 \\
 & 10 & .553 & .069 & .127 & .617 & .219 & .282 & .611 & .224 & .269 & .615 & .227 & .276 & .611 & .116 & .273 & .611 & .216 & .267 \\
 & 25 & .563 & .086 & .125 & .664 & .261 & .332 & .671 & .280 & .324 & .670 & .284 & .320 & .660 & .129 & .319 & .669 & .271 & .320 \\
 & 50 & .557 & .090 & .140 & .664 & .260 & .335 & .705 & .294 & .358 & .708 & .298 & .361 & .664 & .117 & .343 & .698 & .285 & .349 \\
\midrule
\multicolumn{20}{l}{\textit{Reasoning Encoders}} \\
\midrule
\multirow{4}{*}{\textbf{Diver-Retriever}} 
 & 5  & .563 & .098 & .124 & .619 & .209 & .282 & .619 & .210 & .279 & .617 & .213 & .270 & .618 & .110 & .276 & .618 & .199 & .277 \\
 & 10 & .558 & .084 & .104 & .650 & .239 & .325 & .657 & .257 & .323 & .656 & .262 & .318 & .648 & .122 & .319 & .656 & .242 & .323 \\
 & 25 & .582 & .106 & .125 & .666 & .274 & .351 & .681 & .309 & .349 & .681 & .314 & .353 & .663 & .143 & .350 & .684 & .293 & .357 \\
 & 50 & .557 & .118 & .135 & .626 & .285 & .342 & .643 & .304 & .323 & .640 & .309 & .327 & .625 & .147 & .330 & .643 & .291 & .330 \\
\midrule
\multirow{4}{*}{\textbf{RaDeR}} 
 & 5  & .553 & .052 & .078 & .601 & .136 & .197 & .600 & .139 & .193 & .597 & .142 & .190 & .600 & .068 & .195 & .602 & .133 & .195 \\
 & 10 & .543 & .075 & .083 & .630 & .223 & .267 & .631 & .234 & .256 & .632 & .237 & .257 & .625 & .139 & .248 & .631 & .225 & .258 \\
 & 25 & .570 & .087 & .100 & .646 & .261 & .307 & .668 & .299 & .311 & .670 & .304 & .315 & .658 & .175 & .299 & .665 & .287 & .309 \\
 & 50 & .619 & .093 & .102 & .650 & .280 & .345 & .653 & .315 & .334 & .635 & .322 & .328 & .648 & .180 & .315 & .659 & .303 & .320 \\
\midrule
\multirow{4}{*}{\textbf{ReasonIR}} 
 & 5  & .531 & .003 & .041 & .630 & .142 & .260 & .629 & .142 & .255 & .632 & .161 & .258 & .630 & .040 & .255 & .627 & .103 & .249 \\
 & 10 & .547 & .001 & .052 & .638 & .174 & .281 & .633 & .190 & .275 & .641 & .211 & .295 & .632 & .042 & .262 & .635 & .138 & .263 \\
 & 25 & .544 & .019 & .064 & .638 & .198 & .311 & .652 & .246 & .322 & .653 & .268 & .338 & .641 & .048 & .301 & .642 & .183 & .298 \\
 & 50 & .581 & .025 & .072 & .685 & .211 & .338 & .698 & .270 & .350 & .685 & .291 & .361 & .675 & .048 & .328 & .698 & .206 & .304 \\
\bottomrule
\end{tabular}
}
\caption{Comprehensive zero-cost QPP evaluation across all target cutoffs ($k \in \{5, 10, 25, 50\}$) for the \textbf{TEMPO} benchmark. Predictive power is evaluated using AUROC (AUC), Pearson ($r$), and Spearman ($\rho$).}
\label{tab:massive_appendix_tempo}
\end{table*}

\begin{table}[t]
\centering
\small
\begin{tabular}{lccr}
\toprule
\textbf{Subset} & \textbf{MaxScore} & \textbf{Gap@25} & \textbf{$\Delta$} \\
\midrule
FiQA & 0.643 & 0.761 & +0.118 \\
NFCorpus & 0.749 & 0.736 & -0.013 \\
SciDocs & 0.606 & 0.678 & +0.072 \\
SciFact & 0.756 & 0.816 & +0.060 \\
\midrule
\textit{Macro Avg.} & 0.689 & 0.748 & +0.059 \\
\bottomrule
\end{tabular}
\caption{Per-subset AUROC@25 for MaxScore ($s_1$) and Score Gap ($s_1 - s_{25}$), macro-averaged across 11 retrievers on \textbf{BEIR}. $\Delta$ = Gap@25 $-$ MaxScore. Gap@25 outperforms MaxScore on 3 of 4 subsets. NFCorpus, a known outlier in BEIR with very short queries and an atypical corpus structure, is the single exception ($\Delta = -0.013$).}
\label{tab:per_subset_beir}
\end{table}

\begin{table}[t]
\centering
\small
\begin{tabular}{lccr}
\toprule
\textbf{Subset} & \textbf{MaxScore} & \textbf{Gap@25} & \textbf{$\Delta$} \\
\midrule
Biology & 0.452 & 0.629 & +0.177 \\
AoPS & 0.528 & 0.610 & +0.082 \\
Earth Sci. & 0.595 & 0.617 & +0.022 \\
Economics & 0.512 & 0.537 & +0.025 \\
LeetCode & 0.670 & 0.679 & +0.009 \\
Pony & 0.528 & 0.526 & -0.002 \\
Psychology & 0.511 & 0.613 & +0.102 \\
Robotics & 0.624 & 0.610 & -0.014 \\
StackOverflow & 0.543 & 0.601 & +0.058 \\
Sust.\ Living & 0.618 & 0.666 & +0.048 \\
TheoremQA-Q & 0.523 & 0.770 & +0.247 \\
TheoremQA-T & 0.515 & 0.607 & +0.092 \\
\midrule
\textit{Macro Avg.} & 0.552 & 0.622 & +0.071 \\
\bottomrule
\end{tabular}
\caption{Per-subset AUROC@25 for MaxScore ($s_1$) and Score Gap ($s_1 - s_{25}$), macro-averaged across 11 retrievers on \textbf{BRIGHT}. $\Delta$ = Gap@25 $-$ MaxScore. Gap@25 outperforms MaxScore on 10 of 12 subsets. The two exceptions (Pony, Robotics) show differences below 0.015 and are subsets with structured, domain-specific queries where semantic similarity is more informative.}
\label{tab:per_subset_bright}
\end{table}

\begin{table}[t]
\centering
\small
\begin{tabular}{lccr}
\toprule
\textbf{Subset} & \textbf{MaxScore} & \textbf{Gap@25} & \textbf{$\Delta$} \\
\midrule
Cardano & 0.631 & 0.623 & -0.008 \\
Economics & 0.553 & 0.639 & +0.086 \\
Genealogy & 0.649 & 0.766 & +0.117 \\
History & 0.644 & 0.755 & +0.111 \\
HSM & 0.556 & 0.663 & +0.107 \\
Iota & 0.689 & 0.833 & +0.144 \\
Law & 0.404 & 0.511 & +0.107 \\
Monero & 0.516 & 0.584 & +0.068 \\
Politics & 0.606 & 0.728 & +0.122 \\
Quant & 0.615 & 0.560 & -0.055 \\
Travel & 0.479 & 0.608 & +0.129 \\
Workplace & 0.662 & 0.670 & +0.008 \\
\midrule
\textit{Macro Avg.} & 0.584 & 0.662 & +0.078 \\
\bottomrule
\end{tabular}
\caption{Per-subset AUROC@25 for MaxScore ($s_1$) and Score Gap ($s_1 - s_{25}$), macro-averaged across 11 retrievers on \textbf{TEMPO}. $\Delta$ = Gap@25 $-$ MaxScore. Gap@25 outperforms MaxScore on 10 of 12 subsets. The two exceptions (Cardano, Quant) show differences of $\leq$0.05; Quant involves specialized financial queries with peaked score distributions even under failed retrieval.}
\label{tab:per_subset_tempo}
\end{table}

\end{document}